\documentclass[preprint,12pt,3p]{singlecol-new}
\usepackage{natbib,stfloats}
\usepackage{mathrsfs}
\usepackage{amsmath,amssymb,amsfonts}
\usepackage{graphicx}
\usepackage{scrtime}
\usepackage{resizegather}
\usepackage{float}
\usepackage[dont-mess-around]{fnpct}
\usepackage{etoolbox}
\newcommand*{\affmark}[1][*]{\textsuperscript{#1}}
\usepackage{textcomp}
\usepackage{xcolor}

\usepackage{enumitem}
\usepackage{epsfig}
\usepackage{caption}  
\usepackage{multirow}
\usepackage{amsfonts}
\usepackage{subcaption}

\usepackage{footnote}
\usepackage{algorithmicx}
\usepackage{booktabs}
\usepackage{siunitx}
\usepackage{multicol}

\newcommand\Tstrut{\rule{0pt}{2.4ex}}         
\newcommand\Bstrut{\rule[-0.9ex]{0pt}{0pt}}
\def\BibTeX{{\rm B\kern-.05em{\sc i\kern-.025em b}\kern-.08em
		T\kern-.1667em\lower.7ex\hbox{E}\kern-.125emX}}

\makeatletter
\def\theequation{\arabic{equation}}

\makeatother

\begin{document}%

\setcounter{page}{1}

\LRH{S. M. U. Hashmi  et~al.}

\RRH{Implementation of Symbol Timing Recovery for Estimation of Clock Skew}

\VOL{x}

\ISSUE{x}

\PUBYEAR{201X}

\BottomCatch

\CLline

\subtitle{}

\title{Implementation of Symbol Timing Recovery for Estimation of Clock Skew}

\authorA{S.M. Usman Hashmi \protect\affmark[1], Muntazir Hussain \protect\affmark[1],  Fahad Bin Muslim \protect\affmark[1]}
\affA{\affmark[1] Department of Electronics Engineering\\ Iqra University, Islamabad, Pakistan \\
 E-mail: usman16288@yahoo.com, \\
  E-mail: muntazir\_hussain14@yahoo.com,  \\
   E-mail: fahadbinmuslim@gmail.com}
\authorB{Kashif Inayat\protect\affmark[2]}
\affB{\affmark[2] Department of Electronics and Computer Engineering,\\ Hongik University, Sejong, Korea \\
E-mail: kashif\_chaudhary@yahoo.com}
%

%
%
%
%
%
%
%
\authorC{Seong Oun Hwang\protect\affmark[3,*]}
\affC{\affmark[3] Department of Computer Engineering,
	\\ Gachon University, Korea,\\
Email: sohwang@gachon.ac.kr\\\affmark[*]{Corresponding author} }


\begin{abstract}
Time synchronization in any distributed network can be achieved by using application layer protocols for time correction. Time synchronization method proposed in this article uses symbol timing recovery at the physical layer to correct application layer clock. This cross layer methodology diminishes the quantity of message trades needed by application layer for time synchronization thus resulting in energy saving. Precision of skew estimate can be increased by using multiple message exchanges. Examination of the cross layer strategy including the simulation results, the experimentation outcomes and mathematical analysis demonstrates that clock skew at physical layer is same as of application layer, which is actually the skew of hardware clock within the node.
\end{abstract}

\KEYWORD{Time synchronization, Symbol timing recovery, Energy constrained distributed networks, Energy efficient clock skew estimation, Frequency offset estimation, Sensor networks, Bayesian estimation}

\REF{to this paper should be made as follows: Hashmi, S. M. U., Hussain, M., Muslim, F. B., Inayat, K., and Hwang, S. O. (2019) ‘Implementation of Symbol Timing Recovery for Estimation of Clock Skew’, {\it Int. J. Internet Technology and Secured Transactions}, Vol. x, No. x, pp.xxx\textendash xxx.}

\begin{bio}
S.M. Usman Hashmi received his PhD degree in the domain of Telecommunication and Networks. He is an Assistant Professor in, Electronics Engineering Department at Iqra University, Islamabad, Pakistan. His research interest includes signal processing, communication, and time synchronization at the physical layer. \vs{9}

\noindent Muntazir Hussain received his PhD degree in the domain of Electrical, Engineering. He is an Assistant Professor in Electronics Engineering, Department at Iqra University, Islamabad, Pakistan. His research interest, includes the controller and anti-windup compensator design, time-delay system, non-linear control system, and stability of the power control system. \vs{8}

\noindent Fahad Bin Muslim received his PhD degree in the domain of Electronics, Engineering. He is an Assistant Professor in the Electronics Engineering Department at Iqra University, Islamabad, Pakistan. His research interest includes signal processing, communication, electronic design automation and methodology with special emphasis on low-power and high-performance computing system architectures.\vs{8}

\noindent Kashif Inayat received his BE degree in Electronic Engineering in 2014 from, Iqra University Islamabad Campus, Pakistan, his MS degree in Electronic and Computer Engineering in 2019 from Hongik University, Korea. He is a researcher at the Information Security and Machine Learning Lab, Hongik, University, Korea. His research interest includes efficient artificial intelligence computing/architectures, neuromorphic systems, application of asymmetric and code-based cryptographic schemes in embedded systems, blockchain and Internet of Things. \vs{8}

\noindent Seong Oun Hwang received his BS degree in Mathematics in 1993 from Seoul National University, his MS degree in Computer and Communications Engineering in 1998 from Pohang University of Science and Technology, and his PhD degree in Computer Science from Korea Advanced Institute of Science and Technology. He worked as a software engineer at LG-CNS Systems, Inc. from 1994 to 1996. He worked as a senior researcher at Electronics and Telecommunications Research Institute (ETRI) from 1998 to 2007. He worked as a professor with the Department of Software and Communications Engineering of Hongik University from 2008 to 2019. He is currently a professor of Gachon University and an editor of ETRI Journal. His research interests include cryptography, cyber security and artificial intelligence.\\\\
This paper is a revised and expanded version of a paper entitled, "Clock frequency offset estimation using symbol timing recovery",
presented at  {\it International conference on green and human information technology 2019},  Kuala Lumpur, Malaysia, Jan 16-18, 2019.

\end{bio}

\maketitle

\section{Introduction}\label{Sec:1}
Recent advances in the field of Micro-Electro-Mechanical frameworks (MEMS) have steered the dynamic exploration towards outlining exceptionally distributed, energy optimized, cheap and small unattended sensing gadgets appropriate for sensing, handling and imparting information over distributed wireless networks (\cite{b1,b2,b3,b4,b5}). A Wireless Sensor Network (WSN) comprises of a distributed system of such sensing gadgets (nodes) to sense, team up and course information in a certain environment (\cite{b6,b7}). The fundamental part of a WSN is a base station which acts as a portal to gather information from the sensors, and transfer it to a back-end server for extraction of required information. WSNs are utilized in an extensive variety of various applications like ecological and combat zone observing, localizing interlopers, atomic assault, climate monitoring, vital signs observing, recognition of storm, flood, fire or seismic tremors, observing of soil dampness, processing production plants, traffic and car theft tracking etc. (\cite{b8,b9,b14}). 
Distributed systems require extensive amount of coordination among nodes. WSNs usually deals with sensing of physical quantities and reporting the complete picture with respect to time. Requirements of precision of time varies from application to application (\cite{b18,b19, b20}). Like other wireless distributed systems (\cite{b40,b41,b42}), time synchronization is a real concern in WSNs. Sensor nodes must be impeccably time synchronized with one another with the help of electronic counters (clocks) (\cite{b10,b11,b12}). These clocks may have a phase offset because of the distinction in counter values and a skew (frequency offset) because of the distinction in their rates of counting the quartz crystal oscillations embedded in hardware clock of the node (\cite{b28}). Time synchronization is of extreme significance and is needed at diverse layers while outlining a WSN, which involves both the phase and skew estimation and correction (\cite{b13}). Major evaluation parameters for time synchronization in any WSN are precision, efficiency and lifetime. Clock precision points to the allowed error in a system. Efficiency can be termed as number of processes required to synchronize. Decrease in computational load improves energy efficiency. Energy efficiency can also be improved by reducing the message trades. Lifetime refers to the time before re-synchronization. Expanding the lifetime of synchronization enhances the energy efficiency. This would however, effect the accuracy and the precision of the synchronizing clocks.
At the physical layer, time synchronization incorporates clock signal estimation adjusted both in phase and frequency with the transmission node clock using symbol timing recovery methods (\cite{b13, b14}). Feedback and feed-forward systems are available in literature for physical layer symbol timing recovery (\cite{b11, b16, b17, b29, b31,b32,b33}). Feedback systems are more popular as they uses a close loop system and update the time skew using error signal. Feedback timing recovery procedure gives a better estimate of skew. 
At the application layer, time synchronization is vital for the system in order to get rid of timing ambiguity. There are several ways to adjust for both phase and skew (\cite{b10, b18,b19,b20, b21}). Phase offset is computed with message trades whereas conventional methodology for skew estimation is to use regression techniques on the outcomes of few message trades. Many ways are explored by researchers to increase precision of skew estimate by decreasing number of message trades.
The main contribution of this article is that how physical layer symbol timing recovery gives a skew estimate which can correct application layer clock skew. Increases in precision of skew estimate is also possible by increasing message exchanges. 
The rest of the manuscript is organized as follows. Section \ref{Sec:2} presents the symbol timing recovery and skew estimation at physical layer. Section \ref{Sec:3} presents protocols for skew estimation at application layer. Section \ref{Sec:4} presents the concept regarding how the physical layer symbol timing recovery process estimates skew and make it applicable for application layer clock. Section \ref{Sec:5} provides simulation results which demonstrates that clock skew at physical layer and application layer is same. Section \ref{Sec:6} consists of the experimental assessment of our proposed work. Section \ref{Sec:7} provides the mathematical analysis which includes energy efficiency estimation, Cramer Rao Lower Bound (CRLB) and Bayesian estimate of the skew. Finally, section \ref{Sec:8} concludes the discussion.
\section{Physical Layer Symbol Timing Recovery and Skew Estimate}\label{Sec:2}
Physical layer symbol timing recovery refers to the extraction of the time phase offset and skew of sender’s clock (\cite{b30,b31, b32}). Two popular synchronization systems are the feed-forward system and the feedback system.

\begin{figure}[!htbp]
	\centering{\includegraphics[width=3.5in]{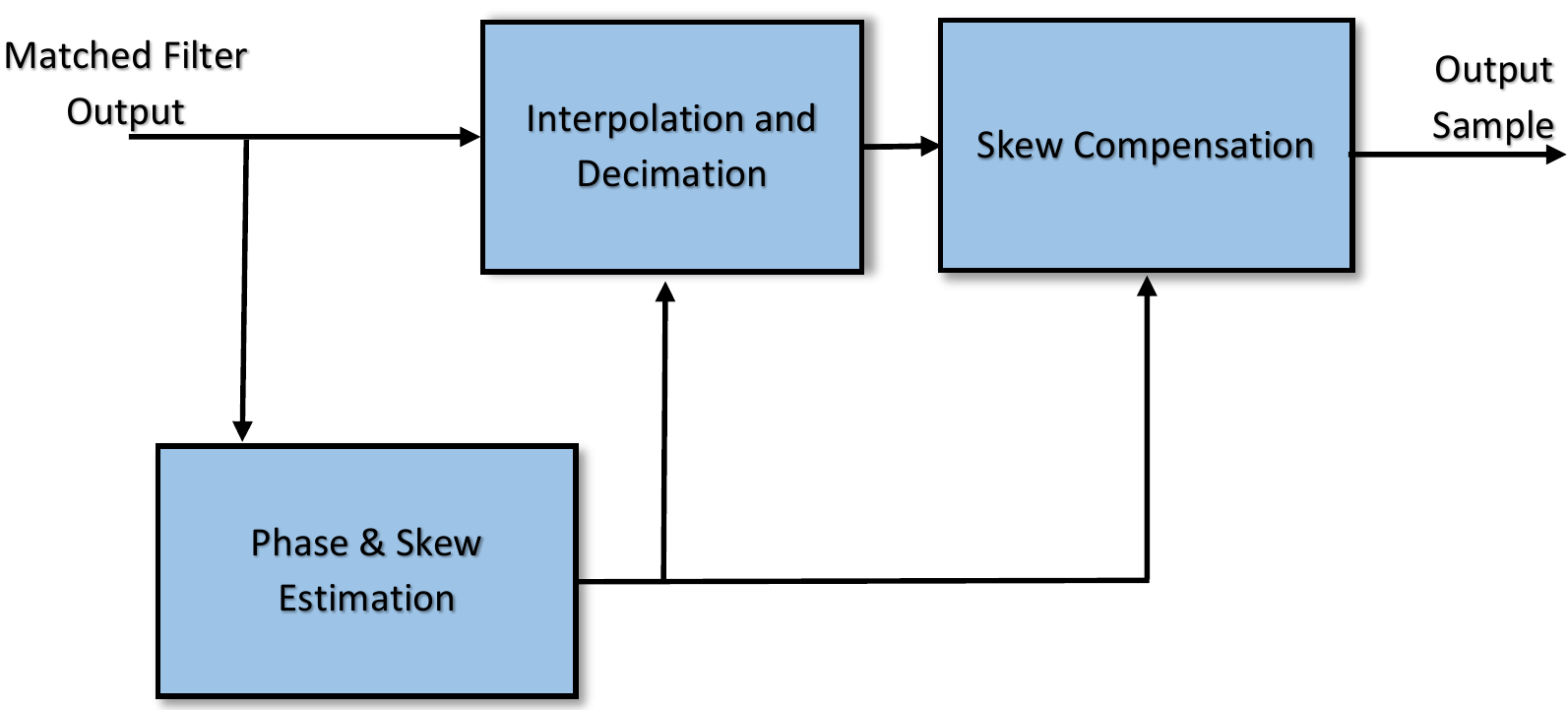}}
	\caption{\textcolor{black}{Feed Forward Symbol Timing Recovery System}.}
	\label{fig1}
\end{figure}

A simple feed forward synchronization system is depicted in Fig. \ref{fig1}. The processing in this method is done sample by sample and the phase and the skew are estimated from the output of the matched filter in a non-recursive manner (\cite{b36,b37,b38}). 

\begin{figure}[!htbp]
	\centering{\includegraphics[width=3.5in]{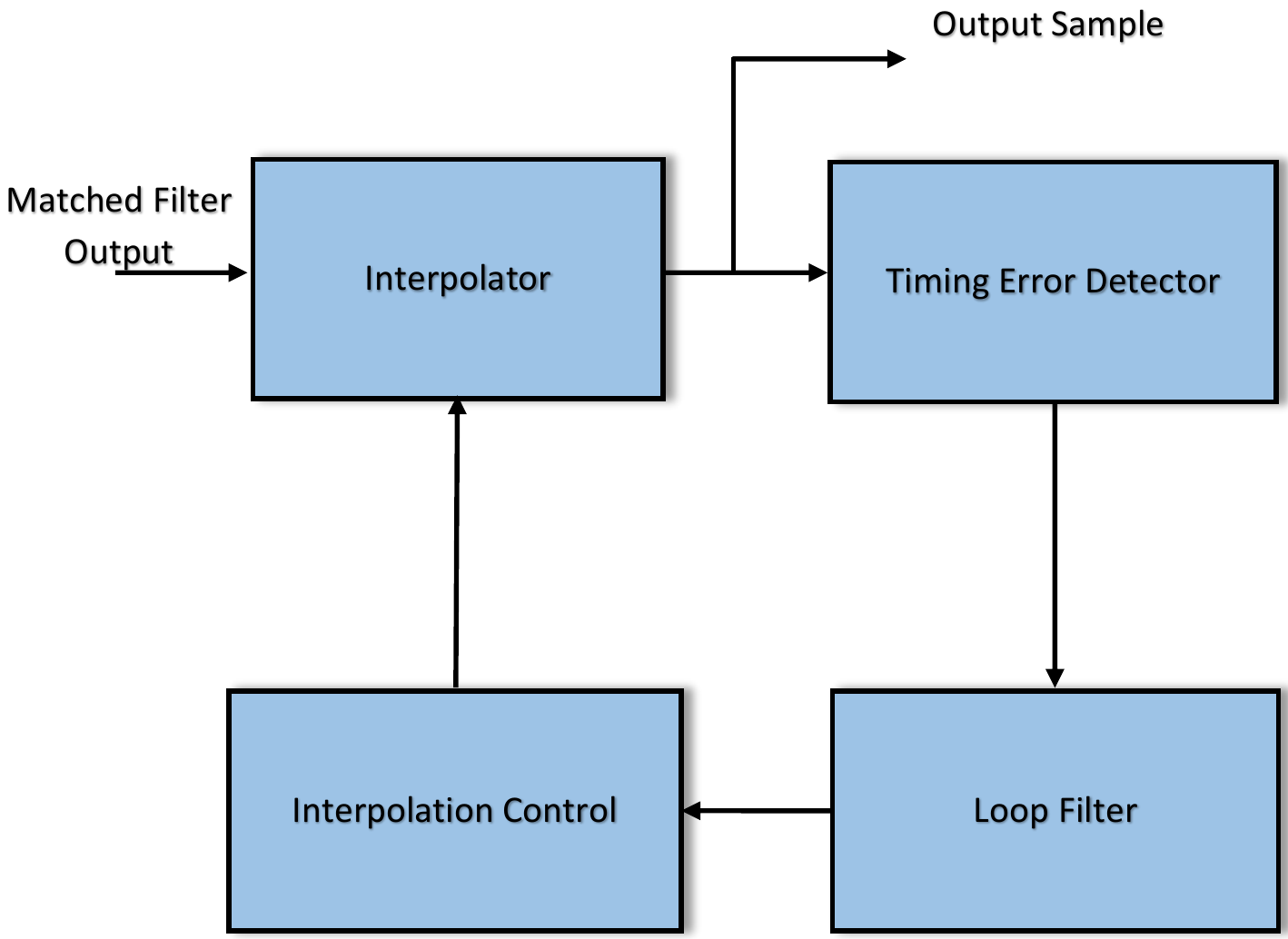}}
	\caption{\textcolor{black}{Feedback Symbol Timing Recovery System}.}
	\label{fig2}
\end{figure}

Feedback systems computes the skew using error in close loop and have more precise estimation than the feed-forward systems (\cite{b30}). The objective is to create discrete samples at the matched filter output during each symbol interval and select the sample that is at an optimum instant. The main components of a feedback system are the interpolator, the timing error detector, the loop filter and the interpolation control, as shown in Fig. \ref{fig2}.
A variety of different timing error detectors (TED) like Zero Crossing TED (ZC-TED), Maximum Likelihood TED (ML-TED) and Gardner TED (G-TED) can find the error signal (\cite{b31,b33}). The idea is to somehow estimate the time error of sample received from previous sample either by using slope, applying sign correction or noticing zero crossing (\cite{b32, b33}). Loop filter tracks the phase and skew of the timing error with the help of error signal provided by TED. Synchronization system performance parameters like acquisition and tracking time are solely dependent on loop filter. A proportional plus integrator loop filter can be a good choice in estimation and tracking of the phase offset as well as skew. Interpolation control can use recursive and non-recursive methods to compute the location of the interpolant which includes sample index as well as distance from last sample (fractional interval). Finally, the interpolant can be computed and placed in an output sample stream with the help of information provided by the interpolation control.
\section{Application Layer Protocols and Skew Estimate}\label{Sec:3}

There are diverse application layer time synchronization protocols available for WSN. Wired network protocols e.g. Network Time Protocol (NTP), Precision Time Protocol (PTP), Time Transmission Protocol (TTP) etc. are not feasible for wireless networks due to their poor energy efficiency (\cite{b18, b19}). Three different time synchronization methods can be implemented as per the requirements of a specific wireless network. Relative timing is by all accounts the least complex one, as it just depends on the sequence of the occasions and do not keep up a real clock. An alternate method for synchronizing is that every node keeps track of its own phase and skew offset estimates and shares this information with every other node in the network. Global synchronization is an alternate system for synchronization which ensures global clock through the whole system. 

\begin{figure}[!htbp]
	\centering{\includegraphics[width=3.5in]{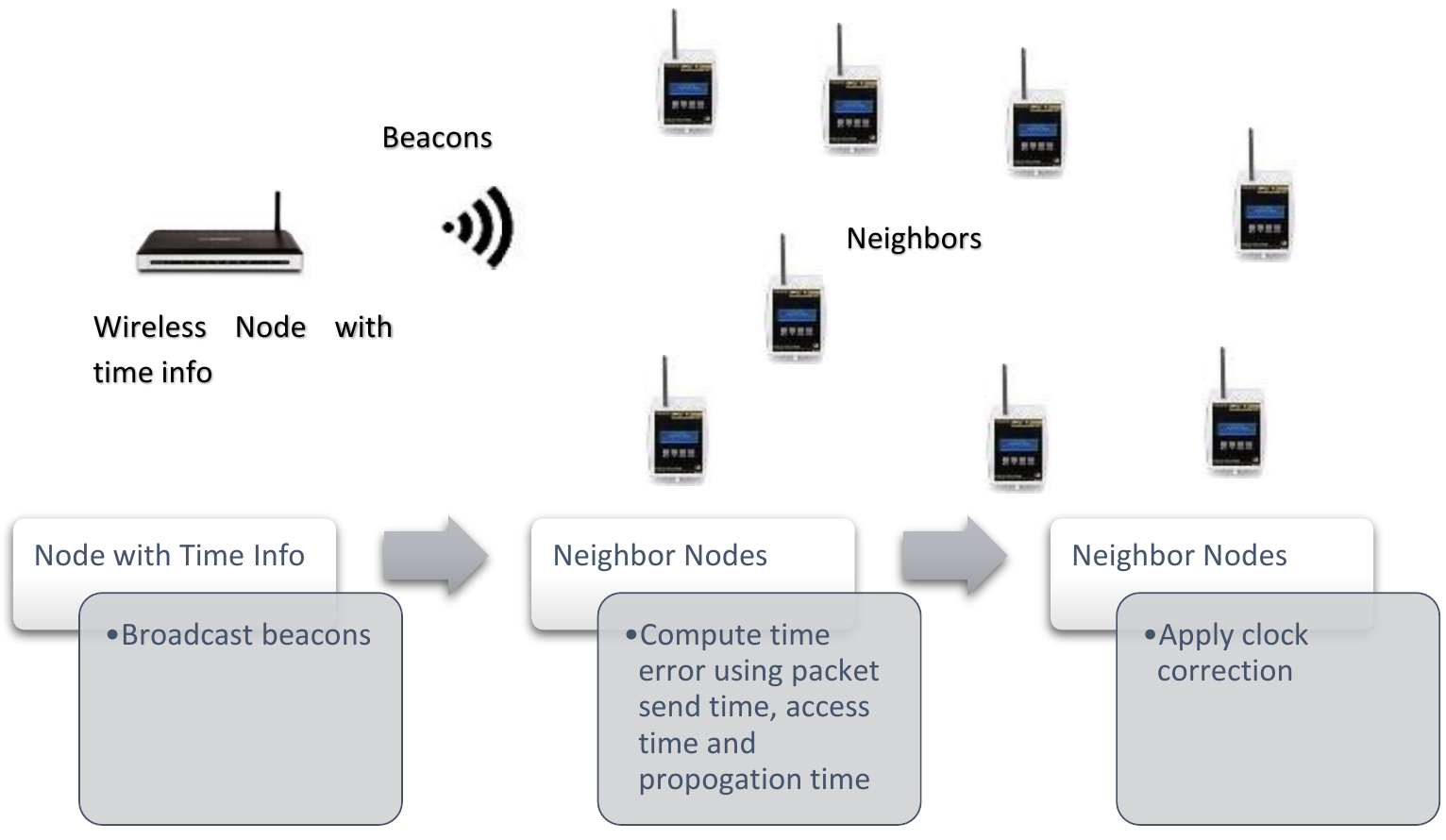}}
	\caption{\textcolor{black}{Reference Broadcast Synchronization Protocol Deployed WSN}.}
	\label{fig3}
\end{figure}

\begin{figure}[!htbp]
	\centering{\includegraphics[width=3.5in]{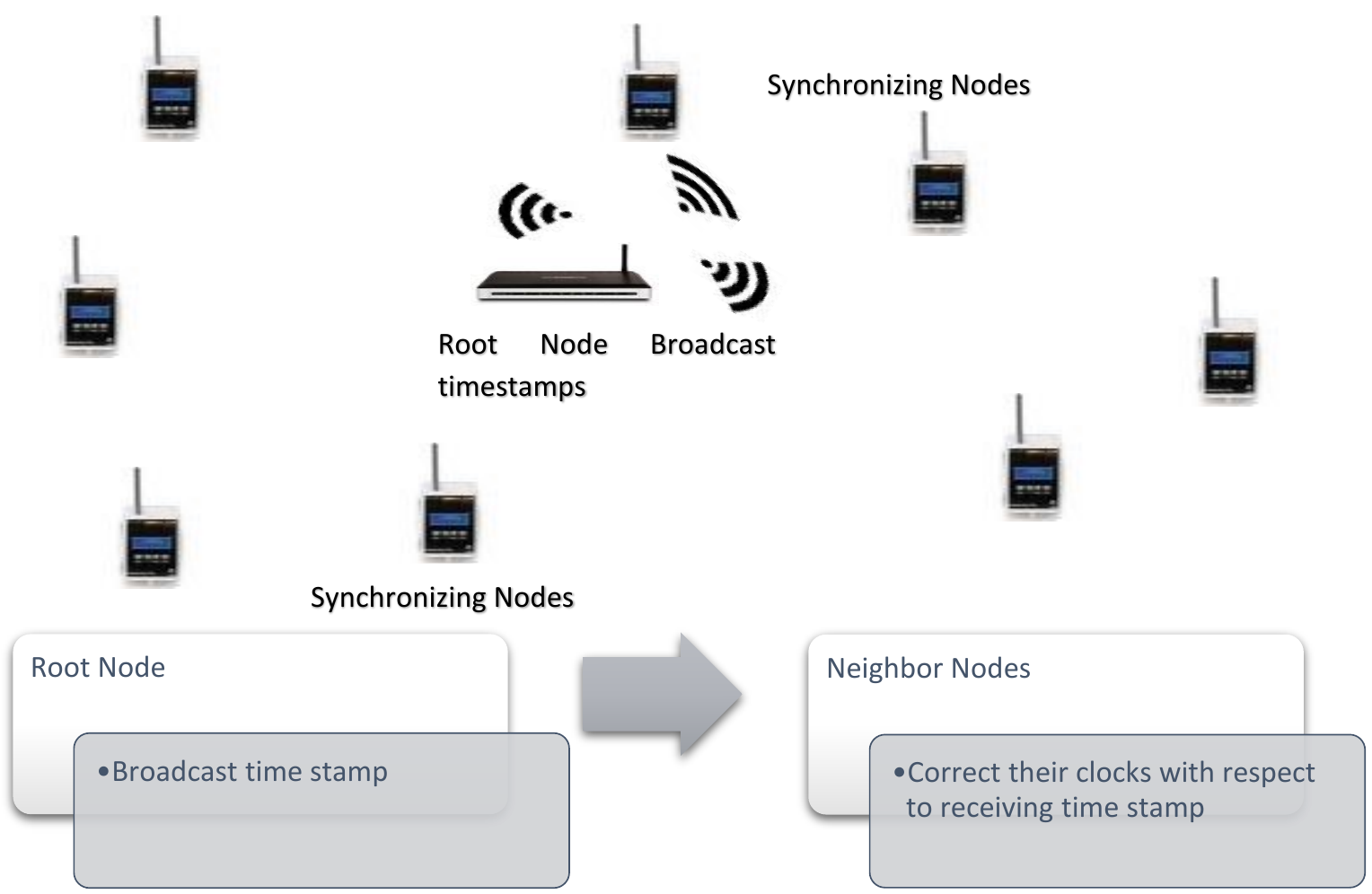}}
	\caption{\textcolor{black}{Flooding time synchronisation protocol deployed WSN}.}
	\label{fig4}
\end{figure}

Reference Broadcast Synchronization (RBS) protocol implements receiver to receiver synchronization by broadcasting reference packet and eventually estimating delays knowing packet send time, access time, propagation time and receive time (\cite{b19, b20, b27}). In RBS, each node maintains a table of phase and skew offset and relates to each other’s local clock without rectification of the clock values. Fig. \ref{fig3} shows the RBS deployed WSN.
Romer’s protocol utilizes message delay to find phase and skew offset between each node and can hold large node density (\cite{b14, b15, b19, b2}). Timing-sync Protocol for Sensor Networks (TPSN) is sender to receiver synchronization protocol that makes tree of the nodes by appointing a level to every node and likewise characterizes the master node. Delay Measurement Time Synchronization, Probabilistic Clock Synchronization, Time-Diffusion Protocol (TDP) are few other protocols with different merits and de-merits. Protocols discussed above and by (\cite{b12, b13, b18, b27, b18}) can be used depending on the requirements of the WSN under consideration. Flooding Time Synchronization Protocol (FTSP) (\cite{b21}) is bandwidth efficient and robust protocol. In FTSP, the master node transmits time packets to every node in the system which correct their clocks by contrasting the sender and receiver clock values. Utilizing FTSP, master node can synchronize in time phase with one packet. However, multiple FTSP packets can estimate skew as well. Fig. \ref{fig4} shows the FTSP deployed in WSN.

\section{Proposed Approach for Skew Estimate}\label{Sec:4}

The skew estimation methodology presented here eliminates the computation and correspondence needed to estimate clock skew at application layer. The idea is to use symbol timing recovery to estimate the clock skew at application layer which shall preserve the energy needed for packetizing and exchanging time stamps. The assembly of a usual wireless sensor node depicts that each process controlled by the processor is based on hardware clock information as shown in Fig 5.
A system model for the proposed approach is shown in Fig. \ref{fig6} which is an extension of our previous work (\cite{b39}). In this revision, the processor utilizes the hardware clock for symbol timing recovery which requires a physical layer clock. The processor also utilizes the same hardware clock to derive the application layer clock of the node.
\begin{figure}[!htbp]
	\centering{\includegraphics[width=3.5in]{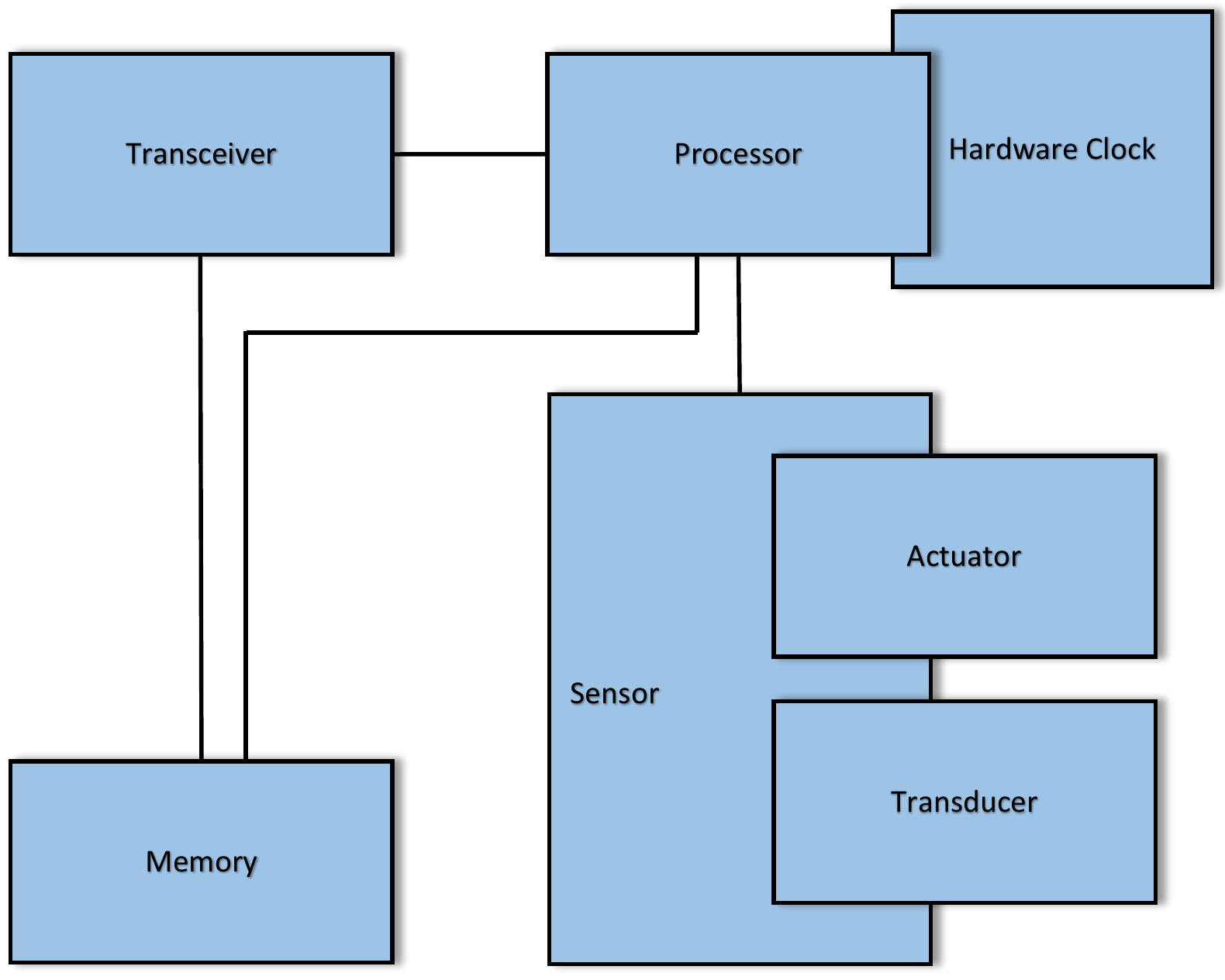}}
	\caption{\textcolor{black}{Architecture of wireless sensor node}.}
	\label{fig5}
\end{figure}
\begin{figure}[!htbp]
	\centering{\includegraphics[width=4in]{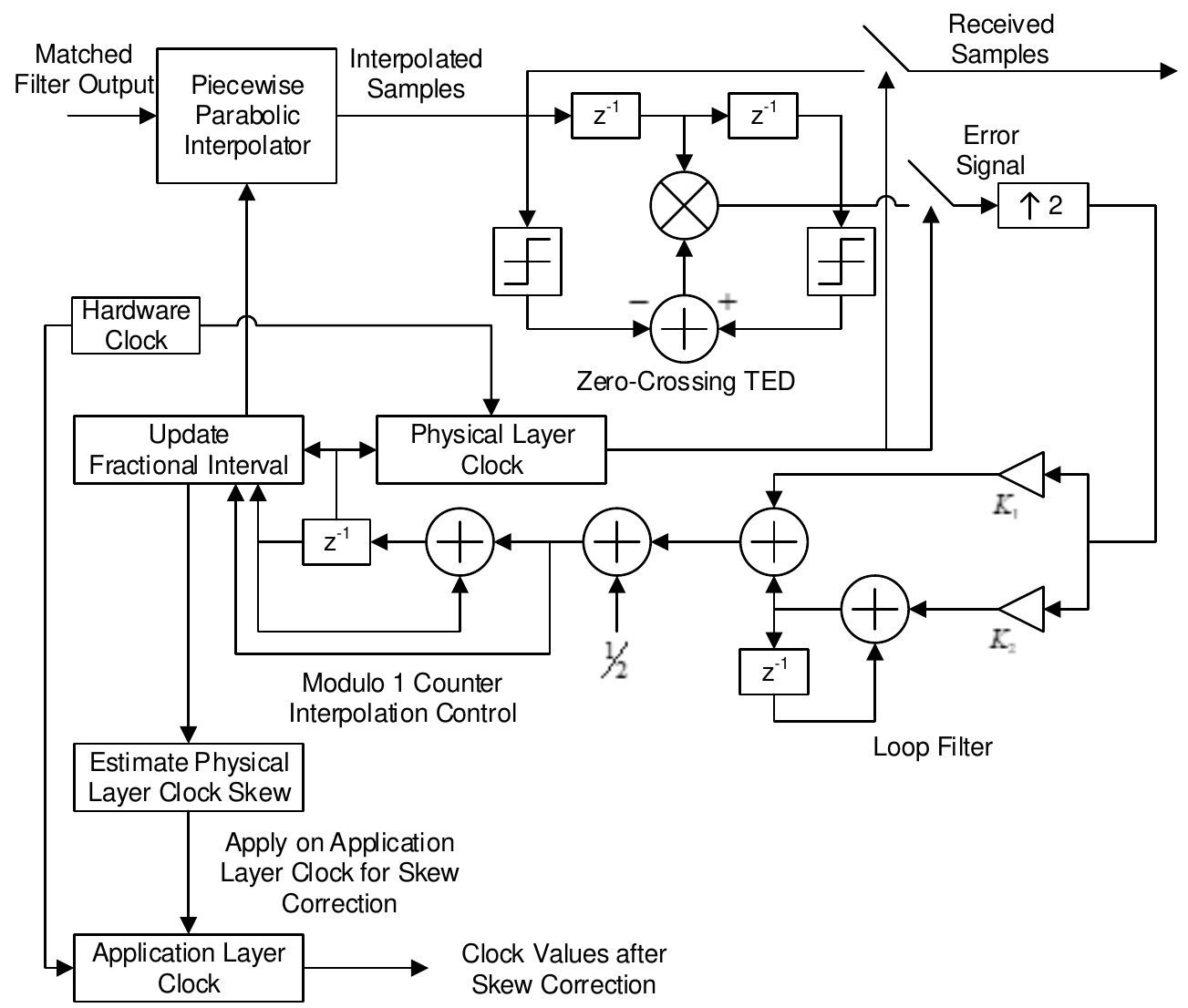}}
	\caption{\textcolor{black}{Proposed system model for application layer clock skew estimation}.}
	\label{fig6}
\end{figure}
This implies that clocks at both layers have the same skew as of hardware clock within one node. The fundamental architecture of the WSN node strengthens the proposed idea of application layer clock skew correction through the clock skew estimation from symbol timing recovery.
The physical layer clock decides when to pick a sample for processing in a WSN node and is derived from the hardware clock. Matched filter’s output at the receiving node must undergo many steps in order to extract the clock skew information. It is a recursive method that continuously estimates and corrects the physical layer clock skew. The loop that tracks and compensates for the skew includes many components shown in Fig 6. Interpolator and ZC-TED works together to find the new sample at optimized location after incorporating error feedback. The error signal from ZC-TED is utilized by a loop filter that has K1 and K2 constants which are carefully adjusted to track phase and skew offset of the physical layer clock. From the output of the loop filter, the interpolation control tries to figure out the base point index (sample location after which a sample needs to be interpolated) and the fractional interval (the time offset after which a sample needs to be interpolated). The fractional interval does have the information of clock skew which can be estimated by any estimation method e.g. least square. Once the clock skew is estimated at the physical layer, it can be applied for clock correction at application layer.

\section{Simulations}\label{Sec:5}
\begin{figure}[!htbp]
	\centering
	\begin{minipage}{.45\linewidth}
		\includegraphics[width=3in]{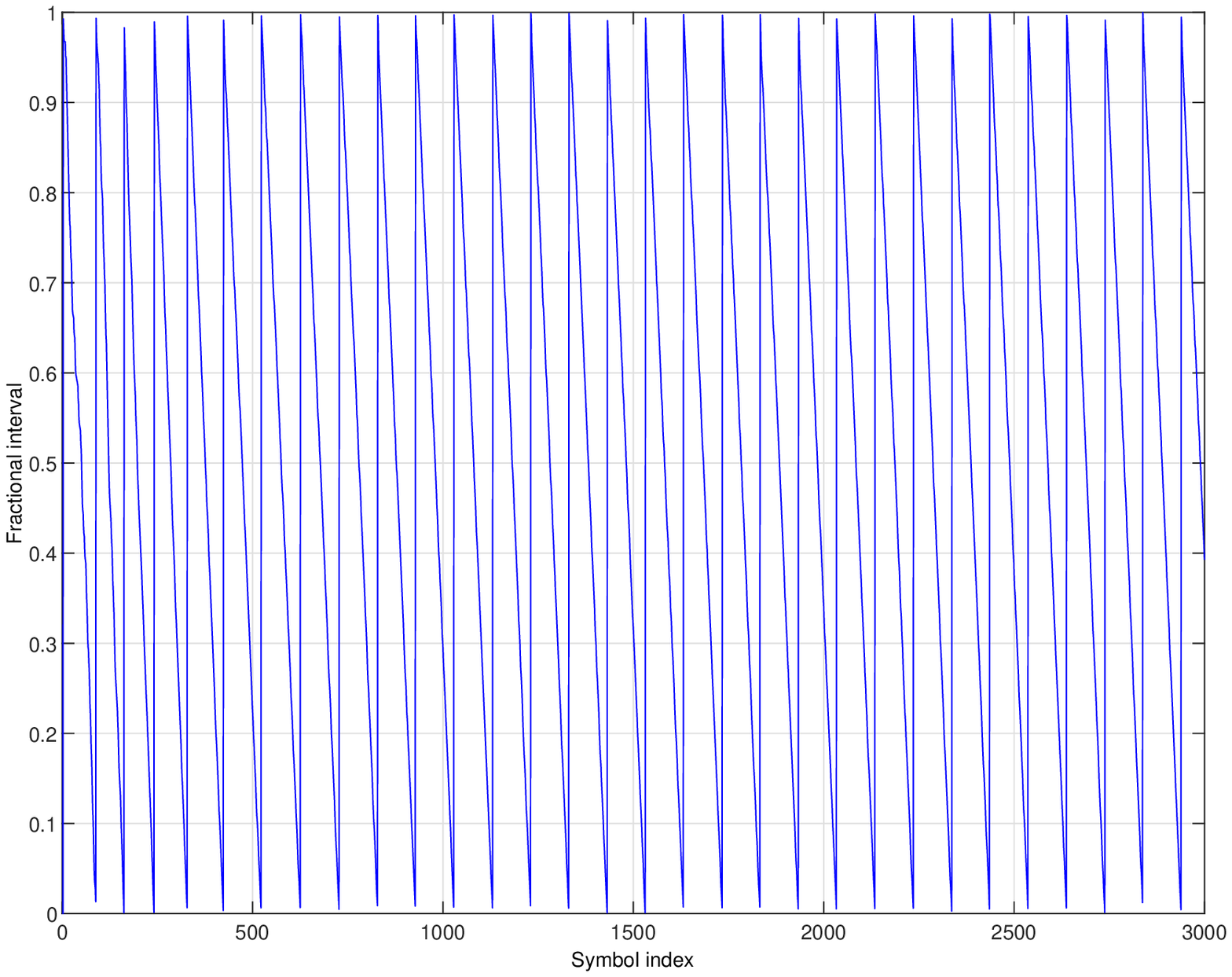}
		\captionof{figure}{Slope of fractional interval shows negative skew of $–4.9505\times10^{–3}$}
		\label{fig7}
	\end{minipage}
	\hspace{.08\linewidth}
	\begin{minipage}{.45\linewidth}
		\includegraphics[width=3in]{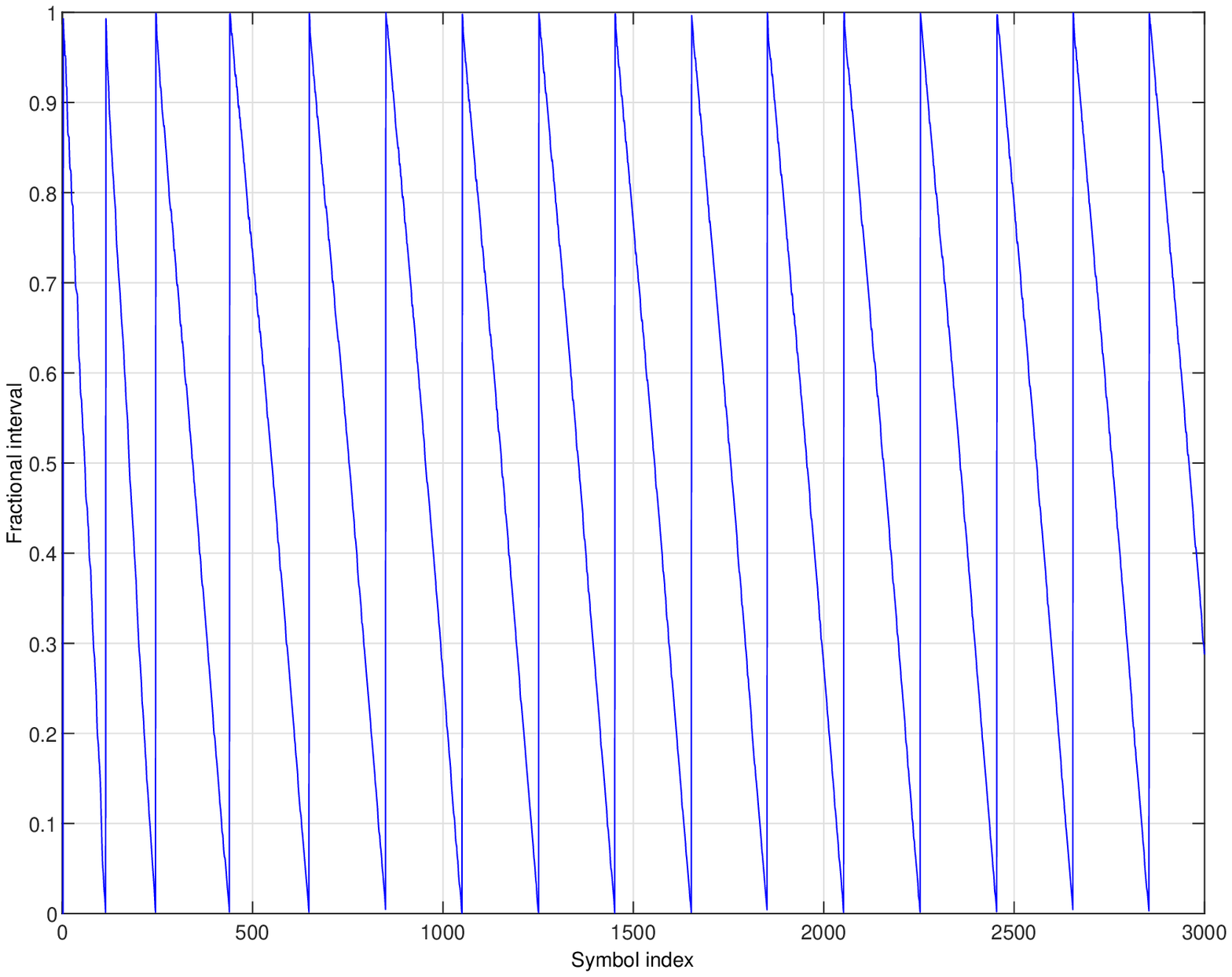}
		\captionof{figure}{Slope of fractional interval shows negative skew of $–2.4938 \times 10^{–3}$}
		\label{fig8}
	\end{minipage}
\end{figure}

Table 1 shows the physical layer configuration utilized in the simulations that is accomplished using the setup proposed in Fig. \ref{fig6}.
\begin{table}[!htbp]
	\centering\scriptsize
	\caption{Physical layer configuration.}
	\smallskip
\begin{tabular}{ll}
	\toprule
	Modulation Scheme & Binary PAM \\
	\hline Symbols & +1,-1 \\
	\hline Total Symbols Generated & 3000 \\
	\hline Symbol Rate & 1000 symbols/sec \\
	\hline Total Transmission Time & $3 \mathrm{sec}$ \\
	\hline Sampling Rate & 8000 samples/sec \\
	\hline Samples per Symbol & 8 \\
	\hline Pulse Shaping Filter & $\mathrm{SRRC}$ \\
	\hline Excess Bandwidth & $50 \%$ \\
	\hline Up-sampling factor & 2\\\bottomrule	
\end{tabular}
\label{Tab:1}
\end{table}
It is assumed that sender and receiver hardware clock skew is $-4.9751\times10^{-3}$. The simulation results in terms of fractional interval, are shown in Fig. \ref{fig7}.

\begin{figure}[!htbp]
	\centering
	\begin{minipage}{.45\linewidth}
		\includegraphics[width=3in]{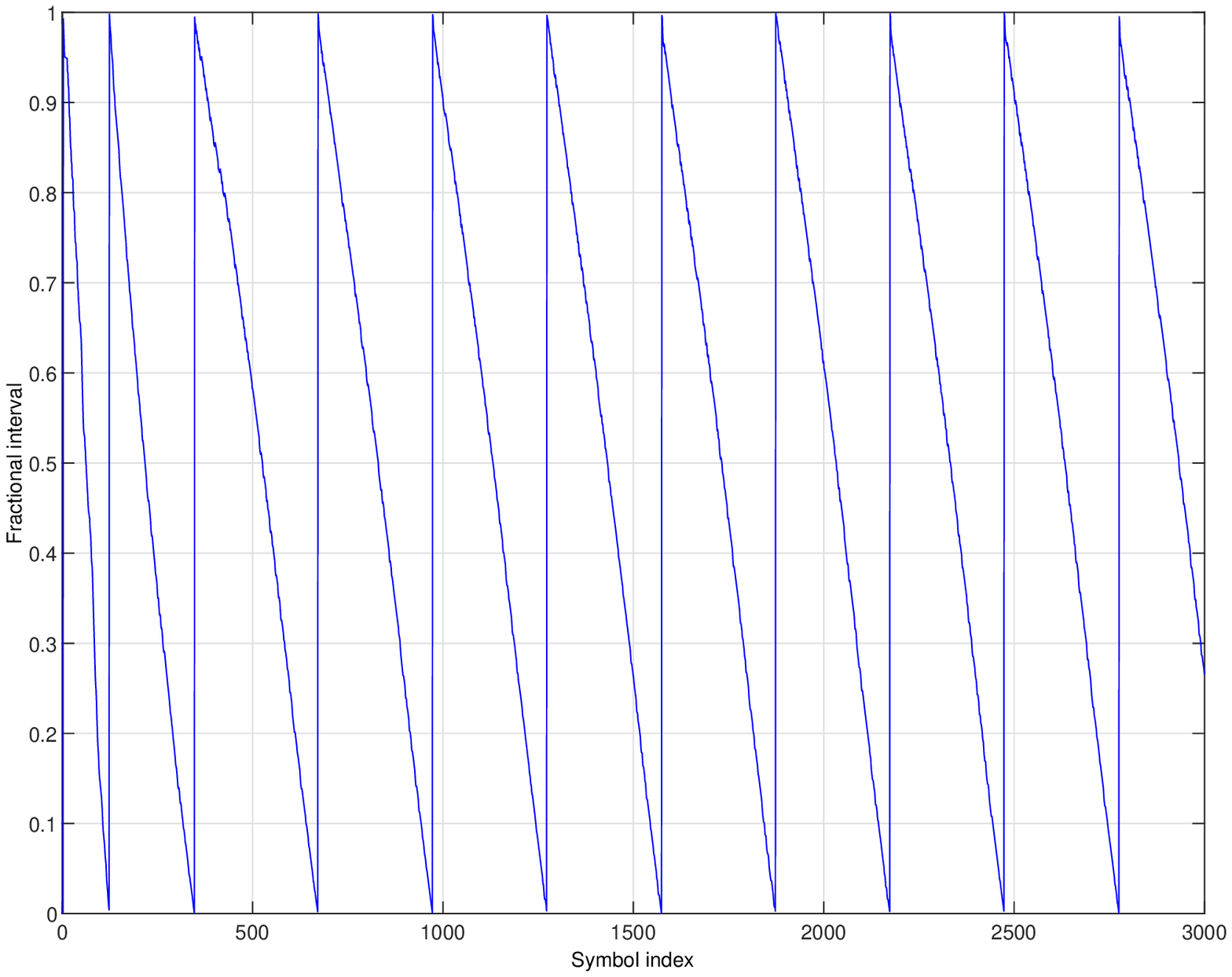}
		\captionof{figure}{Slope of fractional interval shows negative skew of $–1.6639 \times 10^{–3}$}
		\label{fig9}
	\end{minipage}
	\hspace{.08\linewidth}
	\centering
	\begin{minipage}{.45\linewidth}
		\includegraphics[width=3in]{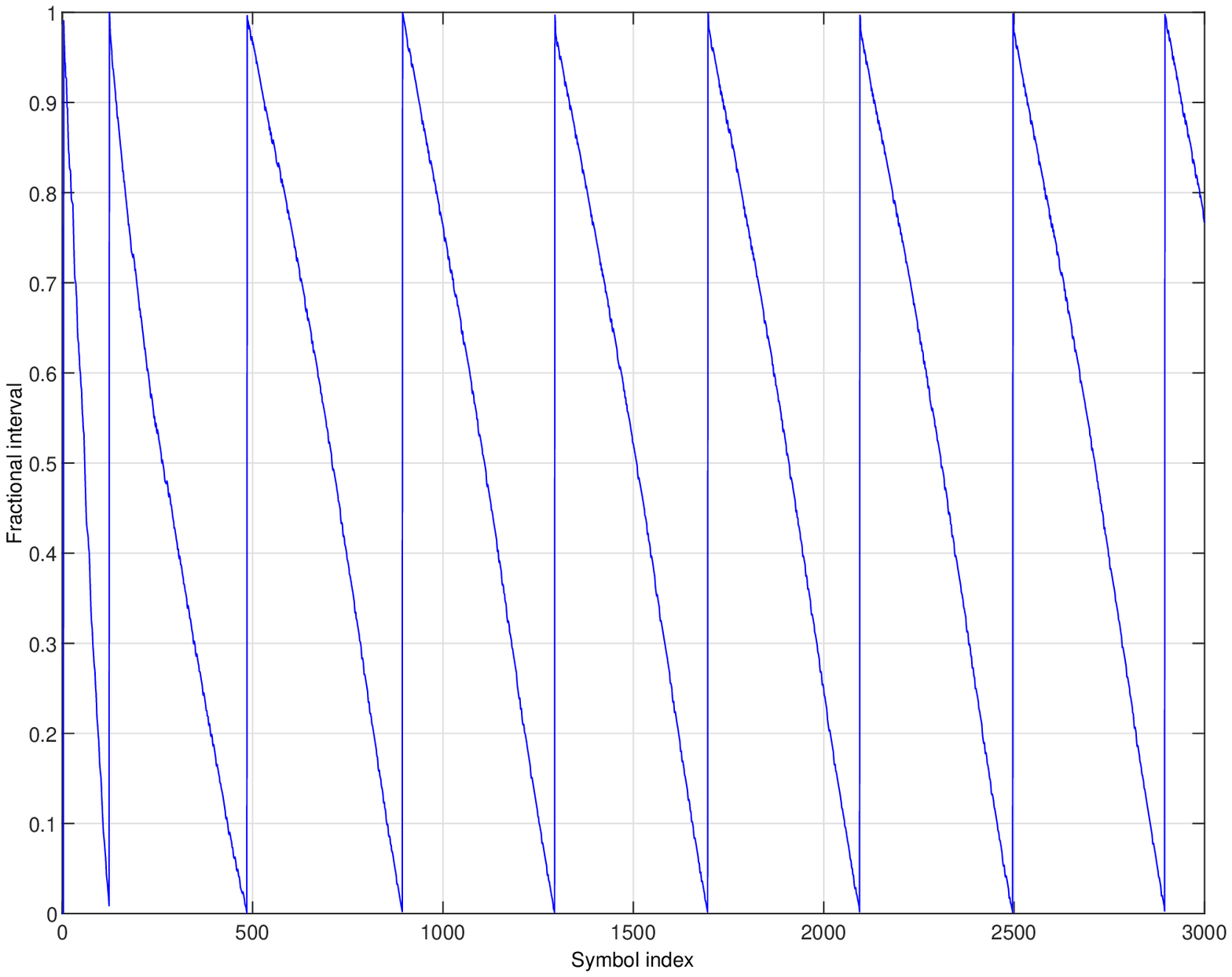}
		\captionof{figure}{Slope of fractional interval shows negative skew of $–1.2484 \times 10^{–3}$}
		\label{fig10}
	\end{minipage}
\end{figure}
\begin{figure}[!htbp]
	\centering
	\begin{minipage}{.45\linewidth}
		\includegraphics[width=3in]{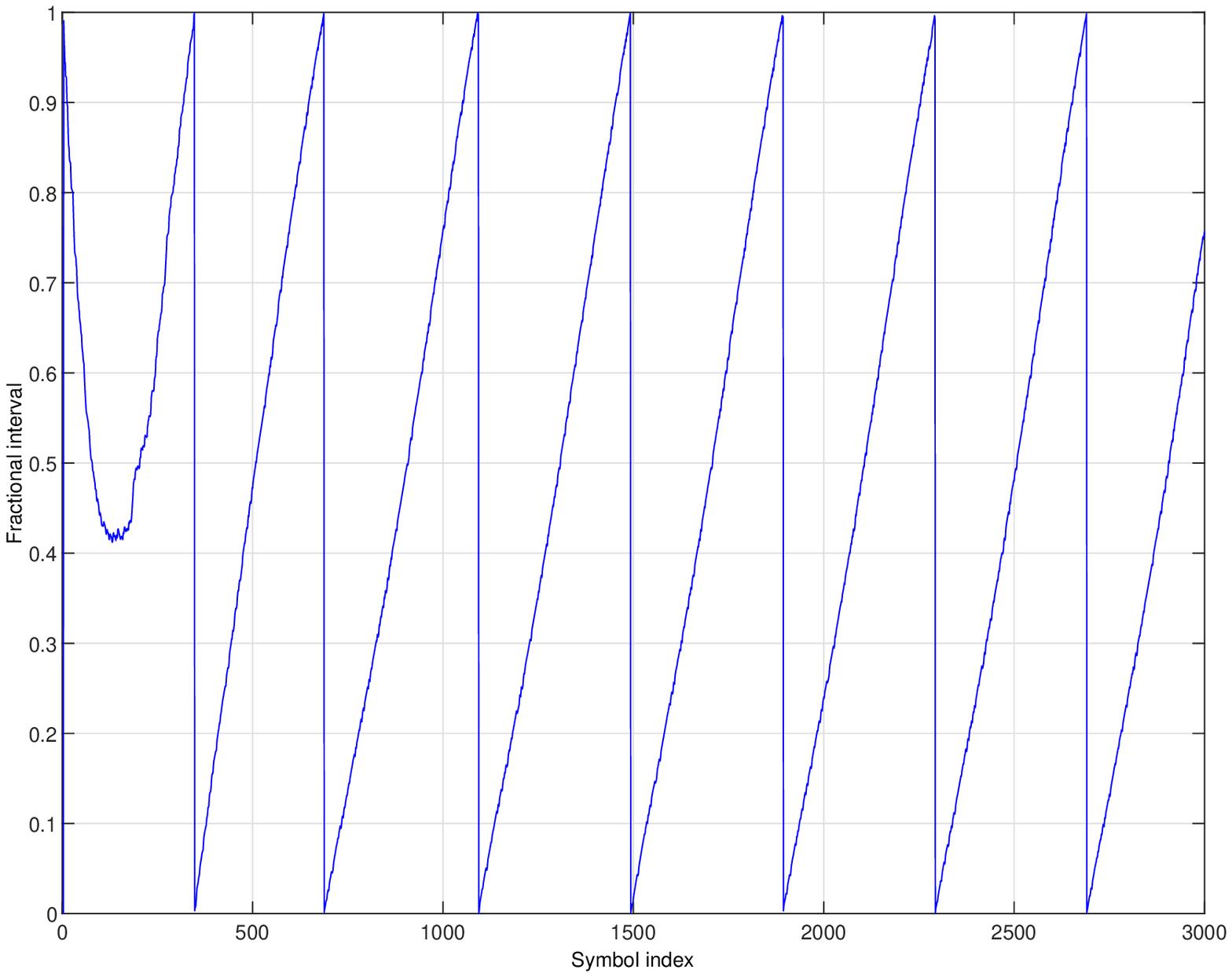}
		\captionof{figure}{Slope of fractional interval shows positive skew of $1.2500 \times 10^{–3}$}
		\label{fig11}
	\end{minipage}
	\hspace{.08\linewidth}
	\centering
	\begin{minipage}{.45\linewidth}
		\includegraphics[width=3in]{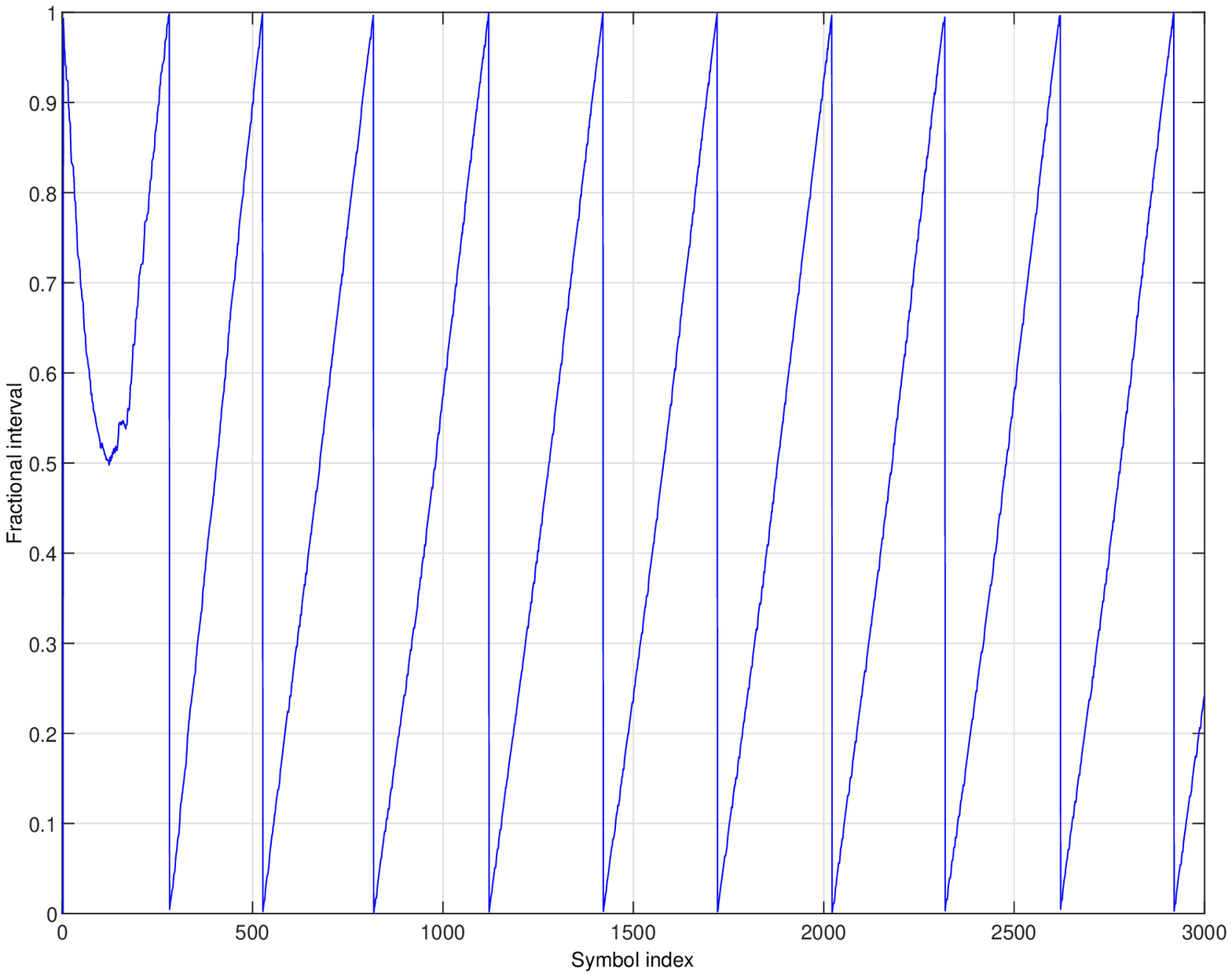}
		\captionof{figure}{Slope of fractional interval shows positive skew of $1.6667 \times 10^{–3}$}
		\label{fig12}
	\end{minipage}
\end{figure}
\begin{figure}[!htbp]
	\centering
	\begin{minipage}{.45\linewidth}
		\includegraphics[width=3in]{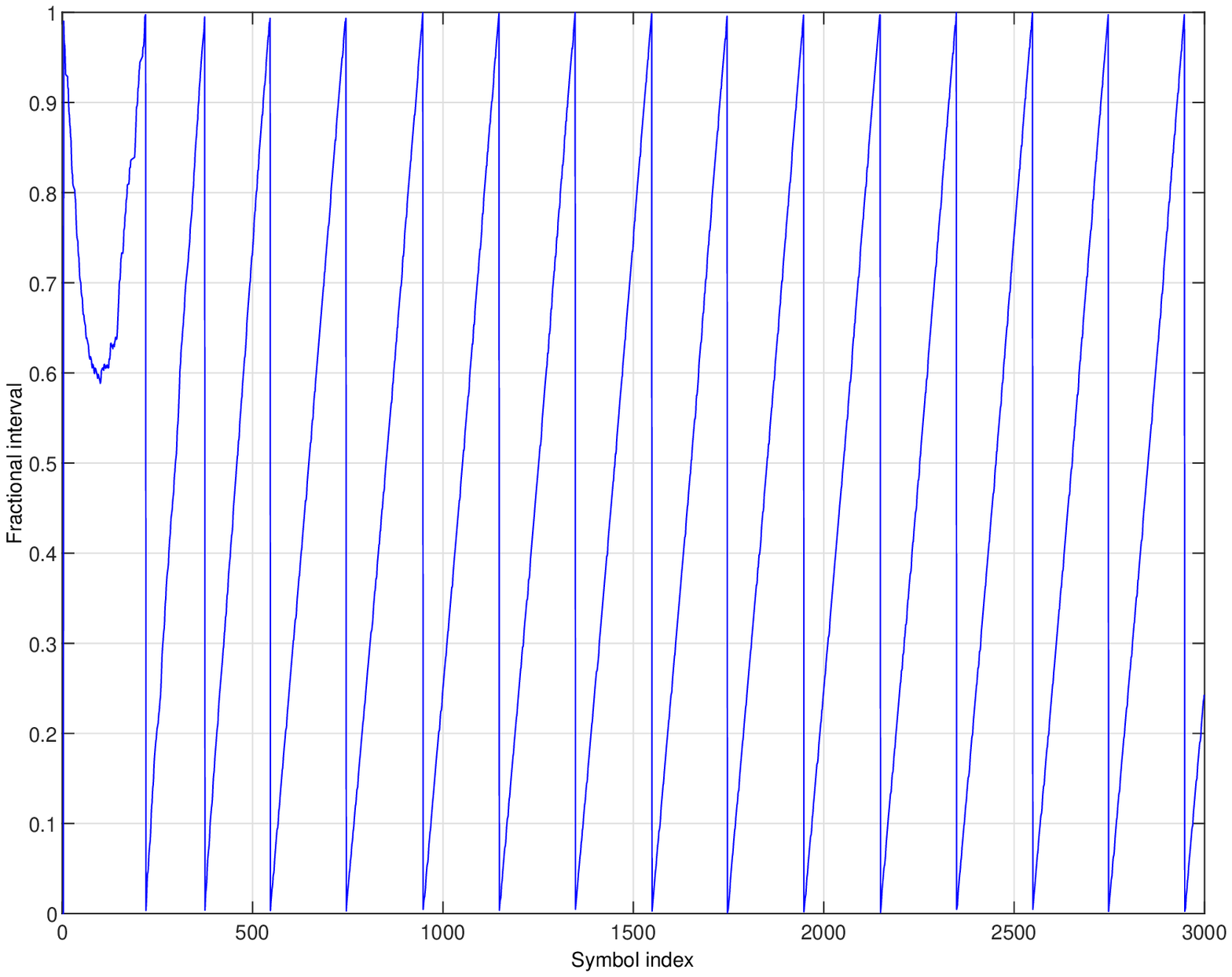}
		\captionof{figure}{Slope of fractional interval shows positive skew of $2.5000 \times 10^{–3}$}
		\label{fig13}
	\end{minipage}
	\hspace{.08\linewidth}
	\centering
	\begin{minipage}{.45\linewidth}
		\includegraphics[width=3in]{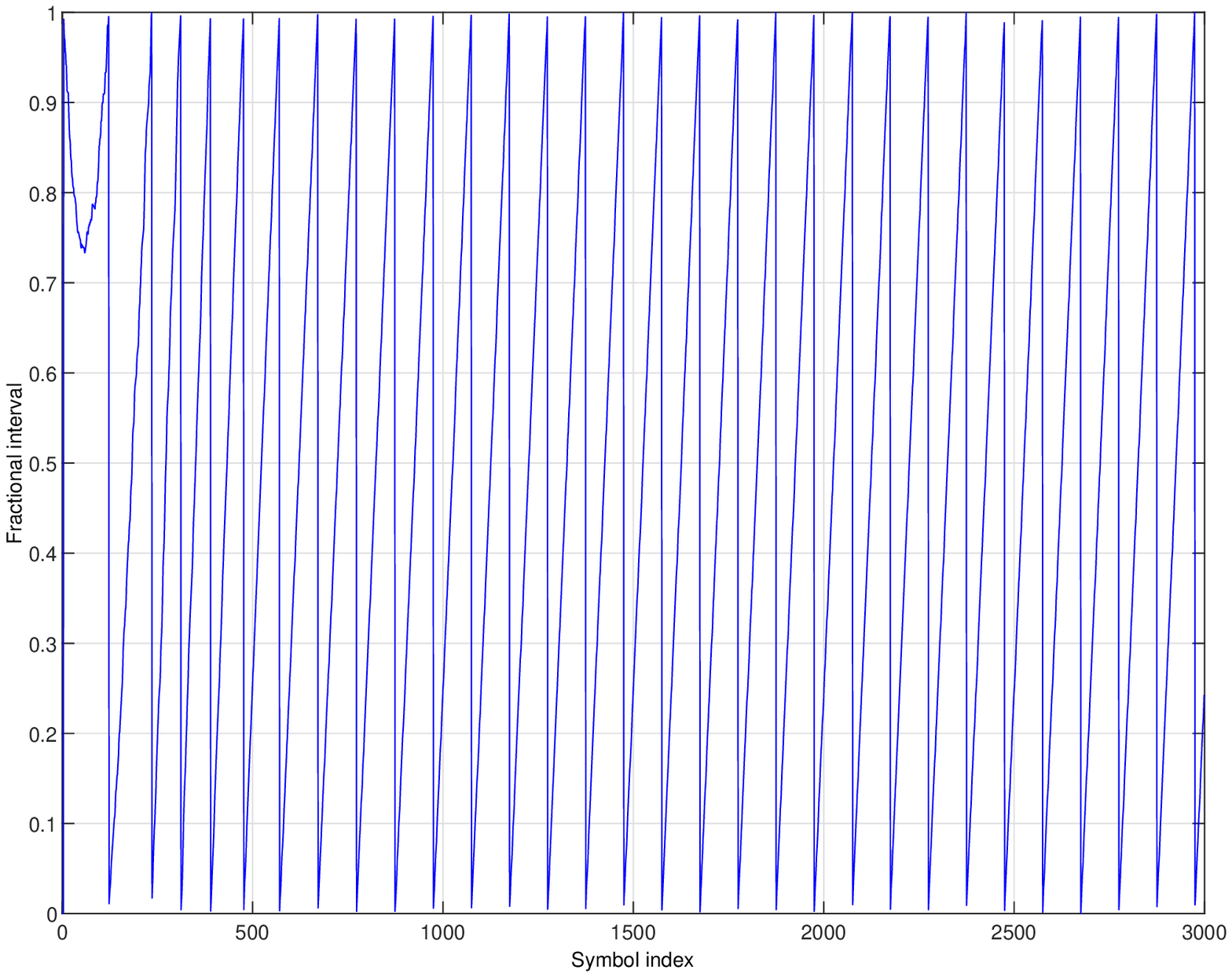}
		\captionof{figure}{Slope of fractional interval shows positive skew of $5.0000 \times 10^{–3}$}
		\label{fig14}
	\end{minipage}
\end{figure}
The skew computed from the simulation equals the slope $-1/101 = -9.90099\times10^{-3}$ divided by the up-sampling factor i.e., $-4.9505\times 10^{-3}$. For the same transmitter receiver pair, we know that the application layer clock skew is $-4.9653\times10^{-3}$ because if the transmitter and receiver clock runs for $500$ seconds, transmitter clock counts from $1$ to $100000$ and receiver clock counts from $1$ to $100499$, then skew is $1-((10000-1)/(100499-1)) = -4.9653\times10^{-3}$. Skew estimated at physical layer is applied to application layer clock of receiver which leaves a negligible error of $0.0148\times10^{-3}$. The data symbols received also include transmitter’s application layer clock phase which can be utilized to correct receiver’s application layer clock phase. After phase and skew correction at application layer, transmitter and receiver clock runs synchronously. Further simulations are carried out with several different transmitter receiver pairs having distinct skew offsets that are summarized below in Table \ref{Tab:2} and Table \ref{Tab:3}.

\begin{table}[!htbp]
	\scriptsize
	\centering
	\caption{Simulation Setup 1, 2, 3, and 4.}
\begin{tabular}{l|llll} 
	\toprule
	\multicolumn{2}{c}{$~~~~~~~~~~~~~~~~~~~~~~$ Simulation I} & Simulation II& Simulation III & Simulation IV\Bstrut\\\hline\hline
	Hardware Skew     & $-4.9751 \times 10^{-3}$ & $-2.4938 \times 10^{-3}$   & $-1.6639 \times 10-3$ & $-1.2484 \times 10-3$     \Tstrut\Bstrut\\
	Tx Clock Values   & 1 to 100000              & 1 to 100000                & 1 to 100000           & 1 to 100000               \\
	Rx Clock Values   & 1 to 100499              & 1 to 100249                & 1 to 100166           & 1 to 100124               \\
	App Layer Skew    & $-4.9653 \times 10^{-3}$ & $-2.4838 \times 10^{-3}$   & $-1.6573 \times 10-3$ & $-1.2385 \times 10-3$     \\
	Phy Layer Skew    & $-4.9505 \times 10^{-3}$ & $-2.4876 \times 10^{-3}$   & $-1.6667 \times 10-3$ & $-1.2438 \times 10-3$     \\
	Error left        & $0.0148 \times 10^{-3}$  & $0.0038 \times 10^{-3}$    & $0.0094 \times 10-3$  & $0.0053 \times 10-3$      \\
	\%age Error       & $0.298 \%$               & $0.153 \%$                 & $0.567 \%$            & $0.428 \%$                 \\
	Simulation Result & Fig. \ref{fig7}                   & Fig. \ref{fig8}                     & Fig. \ref{fig9}                & Fig. \ref{fig10}
	\\\bottomrule
\end{tabular}
	\label{Tab:2}
\end{table}
\begin{table}[!htbp]
	\scriptsize
	\centering
	\caption{Simulation Setup 5, 6, 7, and 8.}
	\begin{tabular}{l|llll} 
		\toprule
		\multicolumn{2}{c}{$~~~~~~~~~~~~~~~~~~~~~~~~~$ Simulation V} & Simulation VI& Simulation VII & Simulation VIII\Bstrut\\\hline\hline
		Hardware Skew     &$1.2500 \times 10^{-3}$ & $1.6667 \times 10^{-3}$& $2.5000 \times 10^{-3}$ & $5.0000 \times 10^{-3}$ \Tstrut\Bstrut\\ 
		Tx Clock Values   &1 to 100000             & 1 to 100000            & 1 to 100000 & 1 to 100000                         \\  
		Rx Clock Values   &1 to 99875              & 1 to 99833             & 1 to 99750 & 1 to 99502                           \\ 
		App Layer Skew    &$1.2516 \times 10^{-3}$ & $1.6728 \times 10^{-3}$& $2.5063 \times 10^{-3}$ & $5.0050 \times 10^{-3}$ \\ 
		Phy Layer Skew    &$1.2469 \times 10^{-3}$ & $1.6667 \times 10^{-3}$& $2.5000 \times 10^{-3}$ & $5.0000 \times 10^{-3}$ \\ 
		Error left        &$0.0047 \times 10^{-3}$ & $0.0061 \times 10^{-3}$& $0.0063 \times 10^{-3}$ & $0.0050 \times 10^{-3}$ \\ 
		\%age Error       &$0.376 \%$              & $0.365 \%$             & $0.251 \%$ & $0.099 \%$                            \\   
		Simulation Result &Fig. \ref{fig11}                 & Fig. \ref{fig12}                & Fig. \ref{fig13} & Fig. \ref{fig14}                                  		\\\bottomrule
	\end{tabular}
	\label{Tab:3}
\end{table}

\section{Experimentations}\label{Sec:6}		
\begin{figure}[!htbp]
	\centering{\includegraphics[width=3.5in]{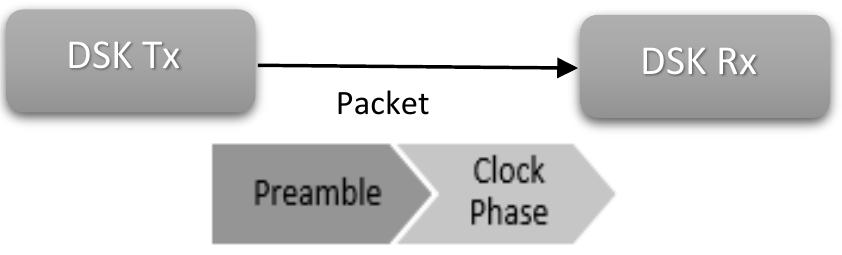}}
	\caption{\textcolor{black}{Experimental setup of WSN with two nodes}.}
	\label{fig15}
\end{figure}

\begin{figure}[!htbp]
	\centering{\includegraphics[width=3in]{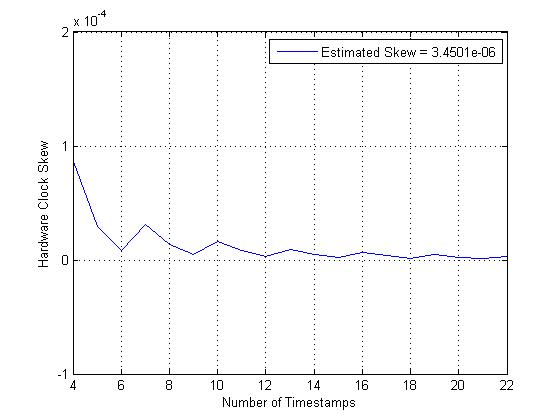}}
	\caption{\textcolor{black}{Hardware clock skew of used DSKs}.}
	\label{fig16}
\end{figure}

A WSN of two nodes is formed with the help of TMS320C6713 DSP Starter Kit (DSK) (\cite{b35}). Each node is modeled by DSK. The DSK transmitter transmits packet to the DSK receiver which includes transmitter’s application layer clock phase as shown in Fig. \ref{fig15}.
The existing skew of hardware clocks of the used DSKs are computed with the help of multiple timestamps and applying regression techniques. Over the exchange of 22 timestamps, the computed skew is of $3.4501\times10^{-6}$ as shown in Fig. \ref{fig16}.

\begin{figure}[!htbp]
	\centering
	\begin{minipage}{.45\linewidth}
		\includegraphics[width=3in]{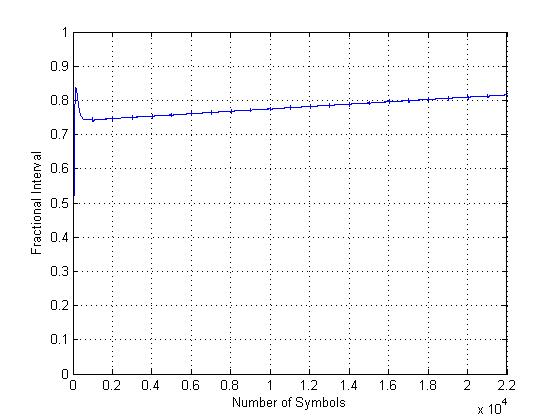}
		\captionof{figure}{Fractional interval with skew of $3.6042 \times 10^{–6}$}
		\label{fig17}
	\end{minipage}
	\hspace{.08\linewidth}
	\begin{minipage}{.45\linewidth}
		\includegraphics[width=3in]{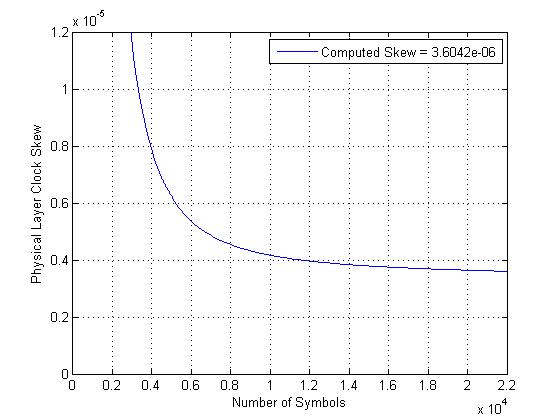}
		\captionof{figure}{Physical layer clock skew estimate for experiment 1.}
		\label{fig18}
	\end{minipage}
\end{figure}

\begin{figure}[!htbp]
	\centering
	\begin{minipage}{.45\linewidth}
		\includegraphics[width=3in]{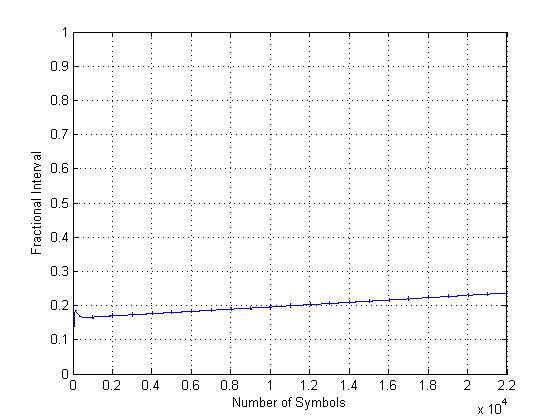}
		\captionof{figure}{Fractional interval with skew of $3.3033 \times 10^{–6}$.}
		\label{fig19}
	\end{minipage}
	\hspace{.08\linewidth}
	\begin{minipage}{.45\linewidth}
		\includegraphics[width=3in]{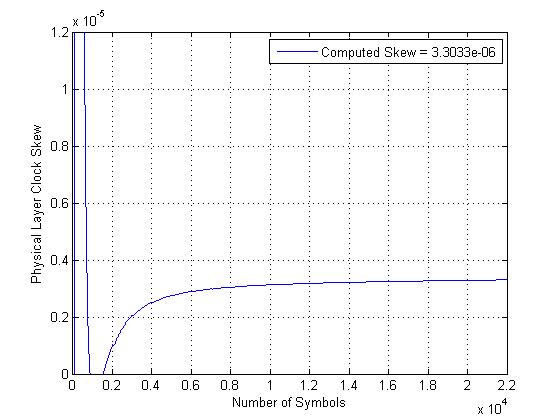}
		\captionof{figure}{Physical layer clock skew estimate for experiment 2.}
		\label{fig20}
	\end{minipage}
\end{figure}

\begin{figure}[!htbp]
	\centering
	\begin{minipage}{.45\linewidth}
		\includegraphics[width=2.9in]{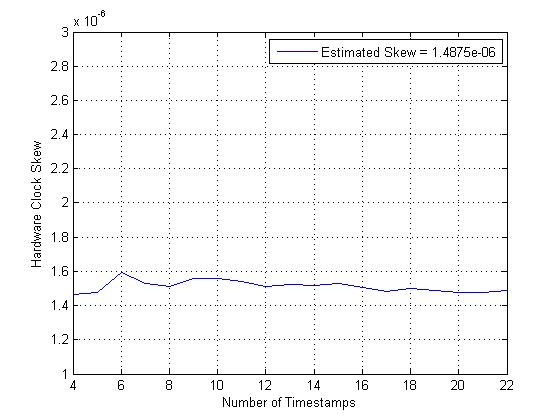}
		\captionof{figure}{Hardware clock skew of used DSKs for experiment 3 and 4.}
		\label{fig21}
	\end{minipage}
	\hspace{.08\linewidth}
	\begin{minipage}{.45\linewidth}
		\includegraphics[width=3in]{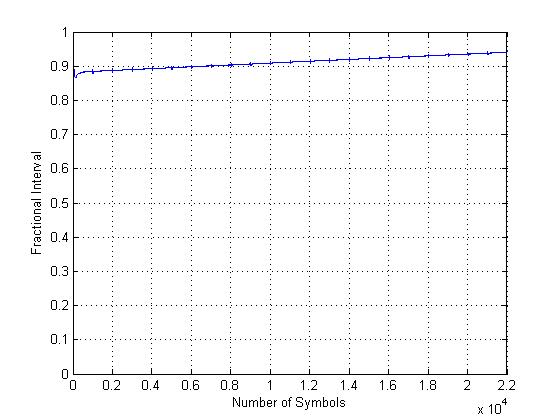}
		\captionof{figure}{Fractional interval with skew of $1.3809 \times 10^{–6}$}
		\label{fig22}
	\end{minipage}
\end{figure}

\begin{figure}[!htbp]
	\centering
	\begin{minipage}{.45\linewidth}
		\includegraphics[width=3in]{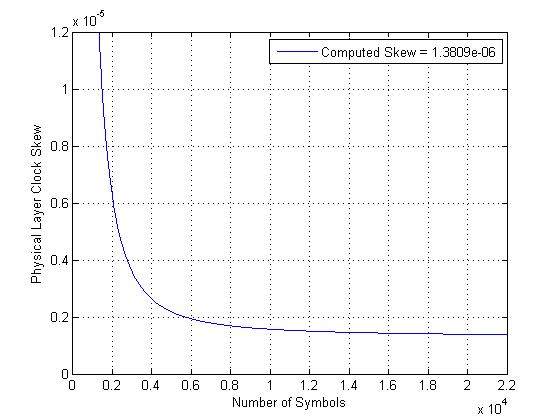}
		\captionof{figure}{Physical layer clock skew estimate for experiment 3}
		\label{fig23}
	\end{minipage}
	\hspace{.08\linewidth}
	\begin{minipage}{.45\linewidth}
		\includegraphics[width=3in]{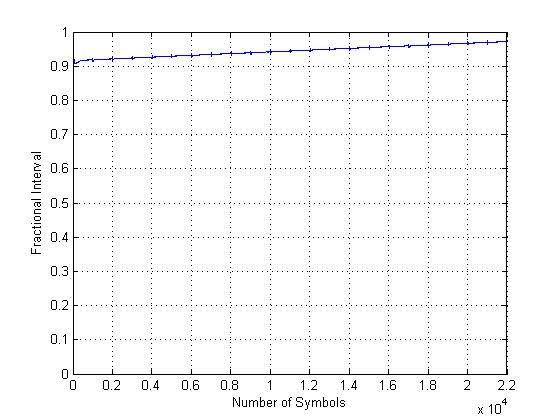}
		\captionof{figure}{Fractional interval with skew of $1.3139 \times 10^{–6}$}
		\label{fig24}
	\end{minipage}
\end{figure}

\begin{figure}[!htbp]
	\centering{\includegraphics[width=3in]{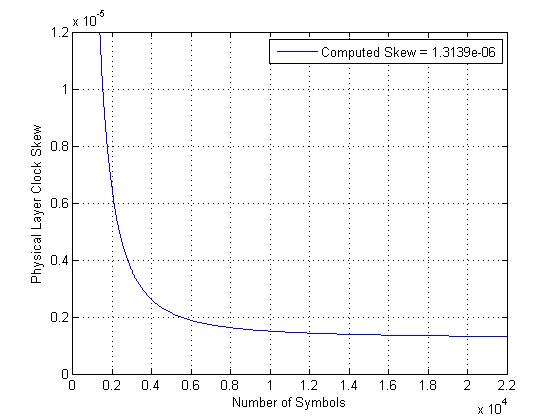}}
	\caption{\textcolor{black}{Physical layer clock skew estimate for experiment 4}.}
	\label{fig25}
\end{figure}

Proposed system model shown in Fig. \ref{fig6} is implemented on DSK Rx which first computes the physical layer clock skew. The physical layer configuration is show in Table \ref{Tab:4}.
\begin{table}[!htbp]
	\centering\scriptsize
	\caption{Physical layer configuration for DSKs.}
	\smallskip	
	\begin{tabular}{ll}
		\toprule
	 Modulation Scheme & Binary PAM \\
	\hline Symbols & +1,-1 \\
	\hline Total Symbols Generated & 22000 \\
	\hline Symbol Rate & 4000 symbols/sec \\
	\hline Total Transmission Time & $5.5 \mathrm{sec}$ \\
	\hline Sampling Rate & 16000 samples/sec \\
	\hline Samples per Symbol & 4 \\
	\hline Pulse Shaping Filter & $\mathrm{SRRC}$ \\
	\hline Excess Bandwidth & $50 \%$ \\
	\hline Up-sampling factor & 2 \\
	\bottomrule
\end{tabular}
\label{Tab:4}
\end{table}
Using fractional interval data (shown in Fig. \ref{fig17}), physical layer clock skew is estimated which comes out to be $3.6042\times10^{-6}$ shown in Fig. \ref{fig18}. This estimated skew is applied for correction of application layer clock skew of DSK Rx that has a hardware skew of $3.4501\times10^{-6}$ with leaves a negligible error of $0.1541\times10^{-6}$. Furthermore the transmitted packet does include the clock phase offset which is utilized to correct application layer clock phase of DSK Rx. Hence one packet is enough to synchronize transmitter-receiver pair in both phase and skew at physical layer as well as application layer.
For 2nd experiment, using the same DSKs yields fractional interval and physical layer clock skew of $3.3033\times10^{-6}$ as shown in Fig. \ref{fig19} and Fig. \ref{fig20}.
3rd and 4th experiment is conducted using a new set of DSKs acting as transmitting and receiving DSKs. The hardware clock skew can be found by repeating the same procedure of exchanging multiple timestamps and it comes out to be $1.4875\times10^{-6}$ as shown in Fig. \ref{fig21}.
The physical layer clock skew computed for 3rd experiment is $1.3809\times10^{-6}$ shown in Fig. \ref{fig22} and Fig. \ref{fig23}, and for the 4th experiment it is $1.3139\times10^{-6}$ shown in Fig. \ref{fig24} and Fig. \ref{fig25}.
All the experimentation results are summarized in Table \ref{Tab:5}:
\begin{table}[!htbp]
	\scriptsize
	\centering
	\caption{Results for experimental setup 1, 2, 3 and 4.}
\begin{tabular}{l|llll} 
	\toprule
	\multicolumn{2}{c}{$~~~~~~~~~~~~~~~~~~~~~~~~~~~~~~~~~$ Experiment  I} & Experiment  II& Experiment  III & Experiment  IV\Bstrut\\\hline\hline
	 Hardware Skew    & $3.4501 \times 10^{-6}$ & $3.4501 \times 10^{-6}$ & $1.4875 \times 10^{-6}$ & $1.4875 \times 10^{-6}$\Tstrut\Bstrut \\
	Phy Layer Skew          & $3.6042 \times 10^{-6}$ & $3.3033 \times 10^{-6}$ & $1.3809 \times 10^{-6}$ & $1.3139 \times 10^{-6}$ \\
	Error left at App Layer & $0.1541 \times 10^{-6}$ & $0.1468 \times 10^{-6}$ & $0.1072 \times 10^{-6}$ & $0.1736 \times 10^{-6}$\\\bottomrule
\end{tabular}
\label{Tab:5}
\end{table}
The error can be further minimized by using more than one packet over the proposed method.
\section{Mathematical Analysis}\label{Sec:7}
Mathematical analysis of the proposed model discussed (All notations used in this section are described in Table \ref{Tab:6})) here includes the minimum possible variance of skew indicated by Cramer-Rao Lower Bound (CRLB), bits required per packet, energy consumption to transmit and receive the packet of proposed method and an application of Bayesian estimate.
Clock skew offset $\tau$ is lower bounded by CRLB (\cite{b31}), can be written as,
\begin{equation}\label{eq1}C R L B(\tau)=\frac{1}{8 \pi^{2} \xi R^{2}} \frac{1}{\frac{E_{S}}{N_{0}}}\end{equation}

Where $\xi=$ loop parameter, $E_S=$  symbol energy and $R =$ symbol rate. From Equation \ref{eq1},

\begin{equation}\label{eq2}R=\sqrt{\frac{1}{8 \pi^{2} \xi C R L B(\tau)} \frac{1}{\frac{E_{\mathrm{S}}}{N_{0}}}}\end{equation}
The number of symbols required to find skew offset $N_S$ can be computed as,

\begin{equation}\label{eq3}N_{S}=0.1125 T_{T} \sqrt{\frac{1}{\xi C R L B(\tau)} \frac{1}{\frac{E_{S}}{N_{0}}}}\end{equation}

Where $T_T$ is transmission time. The packet utilized by the proposed model does include the phase offset and if the symbols required for phase offset are $P_S$ then total symbols $T_S$ can be represented by the following equation,
\begin{equation}\label{eq4}T_{S}=0.1125 T_{T} \sqrt{\frac{1}{\xi C R L B(\tau)} \frac{1}{E_{S}}}+P_{S}\end{equation}

The system utilizes $\alpha$ samples per symbol and $\beta$ bits per sample defines total bits $T_b$ in packet as,

\begin{equation}\label{eq5}T_{b}=\alpha \beta\left(0.1125 T_{T} \sqrt{\frac{1}{\xi C R L B(\tau)} \frac{1}{\frac{E_{S}}{N_{0}}}}+P_{S}\right)\end{equation}
The energy required for transmission of one packet over a distance $x$ with $T_b$ bits can be written as,
\begin{equation}\label{eq6}E_{T x}\left(T_{b}, x\right)=E_{c} T_{b}+\varepsilon T_{b} x^{2}\end{equation}

$E_c$  is the energy required by transceiver circuit and $\varepsilon$  is the gain of the transmitter amplifier. Usual values in case of sensor nodes can be $E_c=50nJm^2$ per bit and $\varepsilon=100pJm^2$ per bit. Similarly, for the reception of same number of bits, energy consumption can be written as,
\begin{equation}\label{eq7}E_{R x}\left(T_{b}\right)=E_{c} T_{b}\end{equation}
The energy consumed by the proposed approach is compared to any application layer skew estimation protocol which at least requires two packets with a very little accuracy. The accuracy of the proposed model i.e. in parts-per-million, can be improved by utilizing multiple packets.
The clock jitter obtained by physical layer clock recovery method is Gaussian (\cite{b30,b31,b34,b39}) and can be seen in simulation and experimentation results and that’s why the probability density function (pdf) for the physical layer skew estimate is Gaussian and can be written as,
\begin{equation}\label{eq8}p(P)=\frac{1}{\sqrt{2 \pi \sigma_{p}^{2}}} e^{-1 / 
	2 \sigma_{p}^{2(p-p p)^{2}}}\end{equation}
Knowing that the skew estimate of physical layer is applied for the correction of application layer clock skew and assuming that $N$ packets of proposed model are used to further increase accuracy, 
\begin{equation}\label{eq9}p(x | P)=\frac{1}{\left(2 \pi \sigma^{2}\right)^{N / 2}} \exp \left[-\frac{1}{2 \sigma^{2}} \sum_{n=0}^{N-1}(x[n]-P)^{2}\right]\end{equation}
Which is the conditional probability density function and $x[n]$ are the packets. If $\bar{x}$ is the mean of application layer clock skew then it can be re-written as follows,
\begin{equation}\label{eq10}p(x | P)=\frac{1}{\left(2 \pi \sigma^{2}\right)^{N / 2}} \exp \left[-\frac{1}{2 \sigma^{2}} \sum_{n=0}^{N-1} x[n]^{2}\right] \exp \left[-\frac{1}{2 \sigma^{2}}\left(N P^{2}-2 N P \bar{x}\right)\right]\end{equation}
Now applying Bayes theorem we have, 
\begin{equation}\label{eq11}p(P | x)=\frac{p(x | P) p(P)}{\int p(x | P) p(P) d P}\end{equation}
Using Equation \ref{eq8} and Equation \ref{eq10}, re-writing above equation results in,
\begin{gather}\label{eq12}p(P | x)=\frac{\frac{1}{\left(2 \pi \sigma^{2}\right)^{N / 2}} \exp \left[-\frac{1}{2 \sigma^{2}} \sum_{n=0}^{N-1} x[n]^{2}\right] \exp \left[-\frac{1}{2 \sigma^{2}}\left(N P^{2}-2 N P \bar{x}\right)\right] \frac{1}{\sqrt{2 \pi \sigma_{P}^{2}}}\exp \left[-1 / 2 \sigma_{{P}}^{2}\left(P-\mu_{P}\right)^{2}\right]}{\int\left(\frac{1}{\left(2 \pi \sigma^{2}\right)^{N / 2}} \exp \left[-\frac{1}{2 \sigma^{2}} \sum_{n=0}^{N-1} x[n]^{2}\right] \exp \left[-\frac{1}{2 \sigma^{2}}\left(N P^{2}-2 N P \bar{x}\right)\right] \frac{1}{\left.\sqrt{2 \pi \sigma_{P}^{2}}\right.} \exp \left[-1 / 2 \sigma_{{P}}^{2}\left(P-\mu_{P}\right)=\right]\right) dP}\end{gather}
Further simplification of the above equation yields,
\begin{gather}\label{eq13}p(P | x)=\frac{\exp \left[-\frac{1}{2}\left(\frac{1}{\sigma^{2}}\left(N P^{2}-2 N P \bar{x}\right)+\frac{1}{\sigma_{P}^{2}}\left(P-\mu_{P}\right)^{2}\right)\right]}{\int\left(\exp \left[-\frac{1}{2}\left(\frac{1}{\sigma^{2}}\left(N P^{2}-2 N P \bar{x}\right)+\frac{1}{\sigma_{P}^{2}}\left(P-\mu_{P}\right)^{2}\right)\right]\right) d P}\end{gather}
And can be written as,
\begin{equation}\label{eq14}p(P | x)=\frac{\exp \left[-\frac{1}{2} Z(P)\right]}{\int\left(\exp \left[-\frac{1}{2} Z(P)\right]\right) d P}\end{equation}
If,
\begin{equation}\label{eq15}Z(P)=\frac{N}{\sigma^{2}} P^{2}-\frac{2 N P \bar{x}}{\sigma^{2}}+\frac{1}{\sigma_{P}^{2}}\left(P-\mu_{P}\right)^{2}\end{equation}
Further simplifying $Z(P)$ gives,
\begin{equation}\label{eq16}Z(P)=\left(\frac{N}{\sigma^{2}}+\frac{1}{\sigma^{2}}\right) P^{2}-2\left(\frac{N}{\sigma^{2}} \bar{x}+\frac{\mu_{P}}{\sigma^{2}}\right) P+\frac{\mu_{P}^{2}}{\sigma^{2}}\end{equation}
Now if the variance of the pdf given by Equation \ref{eq11} is defined by following equation
\begin{equation}\label{eq17}\sigma_{P | x}^{2}=\frac{1}{\frac{N}{\sigma^{2}}+\frac{1}{\sigma_{P}^{2}}}\end{equation}
And the mean can be written as follows,
\begin{equation}\label{eq18}\mu_{P | x}=\left(\frac{N}{\sigma^{2}} \bar{x}+\frac{\mu_{P}}{\sigma^{2}}\right) \sigma_{{P} | x}^{2}\end{equation}
Then Equation \ref{eq16} becomes,
\begin{equation}\label{eq19}Z(P)=\frac{1}{\sigma_{P | x}^{2}}\left(P^{2}-2 \mu_{P | x} P+\mu_{P | x}^{2}\right)-\frac{\mu_{P | x}^{2}}{\sigma_{P | x}^{2}}+\frac{\mu_{P}^{2}}{\sigma_{P}^{2}}\end{equation}
Or further simplification results in,
\begin{equation}\label{eq20}Z(P)=\frac{1}{\sigma_{P | x}^{2}}\left(P-\mu_{P|x}\right)^{2}-\frac{\mu_{P|x}^{2}}{\sigma_{P| x}^{2}}+\frac{\mu_{P}^{2}}{\sigma_{P}^{2}}\end{equation}
Using Equation \ref{eq20}, Equation \ref{eq14} can be written as,
\begin{equation}\label{eq21}p(P | x)=\frac{\exp \left[-\frac{1}{2 \sigma_{P | x}^{2}}\left(P-\mu_{P | x}\right)^{2}\right] \exp \left[-\frac{1}{2}\left(\frac{\mu_{P}^{2}}{\sigma_{P}^{2}}-\frac{\mu_{P | x}^{2}}{\sigma_{P | x}^{2}}\right)\right]}{\int\left(\exp \left[-\frac{1}{2 \sigma_{P | x}^{2}}\left(P-\mu_{P | x}\right)^{2}\right] \exp \left[-\frac{1}{2}\left(\frac{\mu_{p}^{2}}{\sigma_{P}^{2}}-\frac{\mu_{P x}^{2}}{\sigma_{P | x}^{2}}\right)\right]\right) d P}\end{equation}

\begin{equation}\label{eq22}p(P | x)=\frac{1}{\sqrt{2 \pi \sigma_{P| x}^{2}}} \exp \left[-\frac{1}{2 \sigma_{P | x}^{2}}\left(P-\mu_{P | x}\right)^{2}\right]\end{equation}
So posterior pdf comes out to be Gaussian. Now Minimum Mean Square Error (MMSE) can be found by taking the expected value, utilizing Equation \ref{eq17} and \ref{eq18}.
\begin{equation}\label{eq23}\hat{P}=\mu_{P | x}=\frac{\frac{N}{\sigma^{2}} \bar{x}+\frac{\mu_{P}}{\sigma^{2}}}{\frac{N}{\sigma^{2}}+\frac{1}{\sigma^{2}}}\end{equation}
Further simplification results in,
\begin{equation}\label{eq24}\hat{P}=\frac{\sigma_{P}^{2}}{\sigma_{P}^{2}+\sigma^{2} / N} \bar{x}+\frac{\sigma^{2} / N}{\sigma_{P}^{2}+\sigma^{2} / N} \mu_{P}\end{equation}
Now for Bayesian MSE can be written using Equation \ref{eq17} as,
\begin{equation}\label{eq25}B_{m s e}(\hat{P})=\frac{1}{\frac{N}{\sigma^{2}}+\frac{1}{\sigma_{P}^{2}}}\end{equation}
Or,
\begin{equation}\label{eq26}B_{m s e}(\hat{P})=\sigma^{2} / N\left(\frac{\sigma_{P}^{2}}{\sigma_{P}^{2}+\sigma^{2} / N}\right)\end{equation}
If $N_p$ are the total samples and $P[n_P]$ are the samples used to estimate the skew, then mean of the physical layer clock skew can be written as,
\begin{equation}\label{eq27}\mu_{P}=-\frac{6}{N_{P}\left(N_{P}+1\right)} \sum_{n_{P}=0}^{N_{P}-1} P\left[n_{P}\right]+\frac{12}{N_{P}\left(N_{P}^{2}-1\right)} \sum_{n_{P}=0}^{N_{P}-1} n_{P} P\left[n_{P}\right]\end{equation}
And the variance of the physical layer clock skew can be written as,
\begin{equation}\label{eq28}\sigma_{P}^{2}=\frac{1}{N_{P}} \sum_{n_{P}=0}^{N_{P}-1}\left(P\left[n_{P}\right]+\frac{6}{N_{P}\left(N_{P}+1\right)} \sum_{n_{P}=0}^{N_{P}-1} P\left[n_{P}\right]-\frac{12}{N_{P}\left(N_{P}^{2}-1\right)} \sum_{n_{P}=0}^{N_{P}-1} n_{P} P\left[n_{P}\right]\right)^{2}\end{equation}
As defined earlier, $\sigma$ is the variance and $\bar{x}$ is the mean of application layer clock skew which can be written as,
\begin{equation}\label{eq29}\bar{x}=\sum_{m=0}^{N-1} \frac{x[m]}{N}\end{equation}
Using Equation \ref{eq27},\ref{eq28} and \ref{eq29} in Equation \ref{eq24}, we have,
\begin{equation}\label{eq30}\begin{aligned}
&\hat{P}=\frac{\frac{1}{N_{P}} \sum_{n_{P}=0}^{N_{P}-1}\left(P\left[n_{P}\right]+\frac{6}{N_{P}\left(N_{P}+1\right)} \sum_{n_{P}=0}^{N_{P}-1} P\left[n_{P}\right]-\frac{12}{N_{P}\left(N_{P}^{2}-1\right)} \sum_{n_{P}=0}^{N_{P}-1} n_{P} P\left[n_{P}\right]\right)^{2} \sum_{m=0}^{N-1} \frac{x[m]}{N}}{\sigma^{2} /_{N}+\frac{1}{N_{P}} \sum_{n_{P}=0}^{N_{P}-1}\left(P\left[n_{P}\right]+\frac{6}{N_{P}\left(N_{P}+1\right)} \sum_{n_{P}=0}^{N_{P}-1} P\left[n_{P}\right]-\frac{12}{N_{P}\left(N_{P}^{2}-1\right)} \sum_{n_{P}=0}^{N_{P}-1} n_{P} P\left[n_{P}\right]\right)^{2}}\\
&+\frac{\sigma^{2} /_{N}\left(-\frac{6}{N_{p}\left(N_{P}+1\right)} \sum_{n_{P}=0}^{N_{P}-1} P\left[n_{P}\right]+\frac{12}{N_{P}\left(N_{P}^{2}-1\right)} \sum_{n_{P}=0}^{N_{P}-1} n_{P} P\left[n_{P}\right]\right)}{\sigma^{2} /_{N}+\frac{1}{N_{P}} \sum_{n_{P}=0}^{N_{P}-1}\left(P\left[n_{P}\right]+\frac{6}{N_{P}\left(N_{P}+1\right)} \sum_{n_{P}=0}^{N_{P}-1} P\left[n_{P}\right]-\frac{12}{N_{P}\left(N_{P}^{2}-1\right)} \sum_{n_{P}=0}^{N_{P}-1} n_{P} P\left[n_{P}\right]\right)^{2}}
\end{aligned}\end{equation}
And for Bayesian estimate, Equation \ref{eq26} can be re-written as,
\begin{equation}\label{eq31}B_{m se}(\hat{P})=\sigma^{2} / N\left(\frac{\frac{1}{N_{P}} \sum_{n_{P}=0}^{N_{P}-1}\left(P\left[n_{P}\right]+\frac{6}{N_{P}\left(N_{P}+1\right)} \sum_{n_{P}=0}^{N_{P}-1} P\left[n_{P}\right]-\frac{12}{N_{P}\left(N_{P}^{2}-1\right)} \sum_{n_{P}=0}^{N_P-1} n_{P} P\left[n_{P}\right]\right)^{2}}{\sigma^{2} / N+\frac{1}{N_{P}} \sum_{n_{P}=0}^{N_{P}-1}\left(P\left[n_{P}\right]+\frac{6}{N_{P}\left(N_{P}+1\right)} \sum_{n_{P}=0}^{N_{P}-1} P\left[n_{P}\right]-\frac{12}{N_{P}\left(N_{P}^{2}-1\right)} \sum_{n_{P}=0}^{N_{P}-1} n_{P} P\left[n_{P}\right]\right)^{2}}\right)\end{equation}
As proposed in this manuscript, knowing the skew estimate at the physical layer will help us in estimating the application layer clock skew estimate by (\cite{b31}) . Increasing number of cross layer packets will also increase the accuracy of the skew estimate.
\begin{table}[!htbp]
	\scriptsize
	\centering
	\caption{Nomenclature.}
\begin{tabular}{ll} 
	\toprule
	Nomenclature& Description \Bstrut\Bstrut\\
	\hline$\tau$ & Skew offset \\
	$\xi$ & Loop parameter \\
	$E s$ & Symbol energy \\
	$R$ & Symbol rate \\
	$T_{T}$ & Total transmission time \\
	$N_{S}$ & Number of symbols required to find skew offset \\
	$P_{S}$ & Symbols required for phase offset \\
	$T_{S}$ & Total symbols in cross layer packet \\
	$\alpha$ & Samples per symbol \\
	$\beta$ & Bits per sample \\
	$T_{b}$ & Total bits in packet \\
	$E_{r x}$ & Transmit energy \\
	$E_{c}$ & Energy required by transceiver circuit \\
	$x$ & Transmit distance \\
	$p(P)$ & Physical layer skew pdf \\
	$\sigma_{P}^{2}$ & Variance of physical layer skew \\
	$\mu_{P}$ & Mean of physical layer skew \\
	$N$ & Number of cross layer packets \\
	$x[n]$ & Cross layer packets \\
	$\bar{x}$ & Mean of application layer clock skew \\
	$\sigma_{P | x}^{2}$ & Variance of skew offset when physical layer variance is known \\
	$\mu_{P | x}$ & Mean of skew offset when physical layer skew is known \\
	$p(P | x)$ & Posterior pdf of skew offset \\
	$\left|B_{m se}(\hat{P})\right|$ & Bayesian mean square error \\
	$N_{P}$ & Total samples \\
	$P[n_P]$ & Samples used to estimate the skew \\
	\bottomrule
\end{tabular}
\label{Tab:6}
\end{table}
\section{Conclusion}\label{Sec:8}
The proposed method can be utilized to perform synchronization at both physical layer and application layer by using just one packet. The skew estimate of the physical layer is used to correct the application layer clock skew. The phase offset is already there in the sent packet. Hence for a WSN, if each node implements the proposed method, they can synchronize in both phase and skew. Using one packet minimizes the energy consumption in comparison to that needed for computation and transmission of multiple packets by any application layer protocol, however the number of packets can be increased if accuracy needs to be improved in the proposed system. Experimentation and simulation results shows that higher level of accuracy can be achieved with one packet. Mathematical analysis was performed to estimate the number of bits required per packet of the proposed system, the energy required to transmit those bits, CRLB and Bayesian estimate of skew. Future direction of this work may involve the implementation of the model in the case of cooperative communication.



\end{document}